\documentclass{appolb}
\usepackage{epsfig}
\usepackage{graphicx}
\usepackage{amsmath, amssymb}
\usepackage{hyperref}

\newcommand{\mbf}[1]{\mathbf{#1}}

\renewcommand{\bar}[1]{\overline{#1}}

\def\Dslash{\raise.15ex\hbox{/}\kern-.7em D}
\def\Pslash{\raise.15ex\hbox{/}\kern-.7em P}

\begin {document}

\title{
\begin{flushright}
{\small
SLAC-PUB-14258\\
September 2010}
\end{flushright}
QCD  and Light-Front Holography
\thanks{Presented by SJB at the 50th Crakow School, Zakopane. Poland}
}
%
\author {Stanley J. Brodsky\footnote{Electronic address: sjbth@slac.stanford.edu}
\address{SLAC National Accelerator Laboratory\\
Stanford University, Stanford, CA 94309, USA, and \\
CP$^3$-Origins,
Southern Denmark University,
Odense, Denmark}
\and
Guy F. de T\'eramond\footnote{Electronic address: gdt@asterix.crnet.cr} 
{\address{Universidad de Costa Rica, San Jos\'e, Costa Rica}}}
\maketitle
\begin{abstract}

The soft-wall AdS/QCD model, modified by a positive-sign dilaton metric, leads to a remarkable one-parameter description of nonperturbative hadron dynamics. The model predicts a zero-mass pion for zero-mass quarks and a Regge spectrum of linear trajectories with the same slope in the leading orbital angular momentum $L$ of hadrons and the radial  quantum number $N$.  
Light-Front Holography maps the amplitudes which are functions of the fifth dimension variable $z$ of anti-de Sitter space to a corresponding hadron theory quantized on the light front. The resulting Lorentz-invariant relativistic light-front wave equations are functions of  an invariant impact variable $\zeta$ which measures the separation of the quark and gluonic constituents within the hadron at equal light-front time.  The result is 
to a semi-classical frame-independent first approximation to the spectra and light-front wavefunctions of meson and baryon light-quark  bound states,  which in turn predict  the
behavior of the pion and nucleon  form factors.  The theory implements chiral symmetry  in a novel way:    the effects of chiral symmetry breaking increase as one goes toward large interquark separation, consistent with spectroscopic data, and  the 
the hadron eigenstates generally have components with different orbital angular momentum; e.g.,  the proton eigenstate in AdS/QCD with massless quarks has $L=0$ and $L=1$ light-front Fock components with equal probability. The soft-wall model also predicts the form of the non-perturbative effective coupling $\alpha_s^{AdS}(Q)$ and its $\beta$-function
which agrees with the effective
coupling $\alpha_{g_1}$ extracted from  the Bjorken sum rule.
The AdS/QCD model can be systematically improved  by using its complete orthonormal solutions to diagonalize the full QCD light-front Hamiltonian or by applying the Lippmann-Schwinger method in order to systematically include the QCD interaction terms.  A new perspective on quark and gluon condensates is also reviewed.

\end{abstract}



\section{Introduction}

The Schr\"odinger equation plays a central role in atomic physics, providing a simple,  but effective, first approximation description of  the spectrum and wavefunctions of bound states in quantum electrodynamics. It can be systematically improved in QED, leading to an exact quantum field theoretic description of  atomic states such as positronium and muonium as given by the  relativistic Bethe-Salpeter equation.

A long-sought goal in hadron physics is to find a simple analytic first approximation to QCD analogous to the Schr\"odinger-Coulomb equation of atomic physics.	This problem is particularly challenging since the formalism must be relativistic, color-confining, and consistent with chiral symmetry.

We have recently shown that  
the soft-wall AdS/QCD model, modified by a positive-sign dilaton metric, leads to a remarkable one-parameter description of nonperturbative hadron dynamics. The model predicts a zero-mass pion for zero mass quarks and a Regge spectrum of linear trajectories with the same slope in the leading orbital angular momentum $L$ of hadrons and the radial  quantum number $N$.  

Light-Front Holography~\cite{deTeramond:2008ht}  maps the amplitudes which are functions of the fifth dimension variable $z$ of anti-de Sitter (AdS) space to a corresponding hadron theory quantized at fixed light-front time $\tau= x^0 +x^3/c$ , the time marked by the
front of a light wave.~\cite{Dirac:1949cp}  The resulting Lorentz-invariant relativistic light-front (LF) wave equations are functions of  an impact variable $\zeta$ which measures the invariant separation of the quark and gluonic constituents within the hadron at equal light-front time; it is analogous to the radial coordinate $r$ in the Schrodinger equation.
This novel approach
leads to a semi-classical frame-independent first approximation to the spectrum and light-front wavefunctions of meson and baryon light-quark  bound states, which in turn predict  the
behavior of the pion and nucleon  form factors in the space-like and time-like regions.  
The resulting equation for a meson $q \bar q$ bound state  has the form of a relativistic Lorentz invariant  Schr\"odinger equation
\begin{equation} \label{eq:QCDLFWE}
\left(-\frac{d^2}{d\zeta^2}
- \frac{1 - 4L^2}{4\zeta^2} + U(\zeta) \right)
\phi(\zeta) = \mathcal{M}^2 \phi(\zeta),
\end{equation}
where the confining potential is $ U(\zeta) = \kappa^4 \zeta^2 + 2 \kappa^2(L+S-1)$ 
in a soft dilaton modified background.~\cite{deTeramond:2009xk}
There is only one parameter, the mass scale $\kappa \sim 1/2$ GeV, which enters the confinement potential. Here $S=0,1$ is the spin of the $q $ and $ \bar q $.  In a relativistic theory, the hadron eigenstate has components of different orbital angular momentum, just as in the Dirac-Coulomb equation, where the lowest eigenstate has both S and P states. By convention one uses the minimum $L$ to label the state.   Furthermore, in light-front theory, the label $L$ refers to the maximum orbital angular projection $|L^z|$,   of the bound state.

The single-variable LF Schr\"odinger equation,  Eq. \ref{eq:QCDLFWE},
determines the eigenspectrum and the light-front wavefunctions (LFWFs)
of hadrons for general spin and orbital angular momentum.~\cite{deTeramond:2008ht}. This
LF wave equation serves as a semiclassical first approximation to QCD,
and it is equivalent to the equations of motion which describe the
propagation of spin-$J$ modes in  AdS space.  Light-front holography
thus provides a remarkable connection between the description of
hadronic modes in AdS space and the Hamiltonian formulation of QCD in
physical space-time quantized on the light-front  at fixed LF
time $\tau.$  If one further chooses  the constituent rest frame (CRF)~\cite{Danielewicz:1978mk,Karmanov:1979if,Glazek:1983ba}  
where the total 3-momentum vanishes: $\sum^n_{i=1} \mbf{k}_i \! = \! 0$, then the kinetic energy in the LF wave equation displays the usual 3-dimensional rotational invariance.

\begin{figure}[!]
\begin{center}
\includegraphics[angle=0,width=8.5cm]{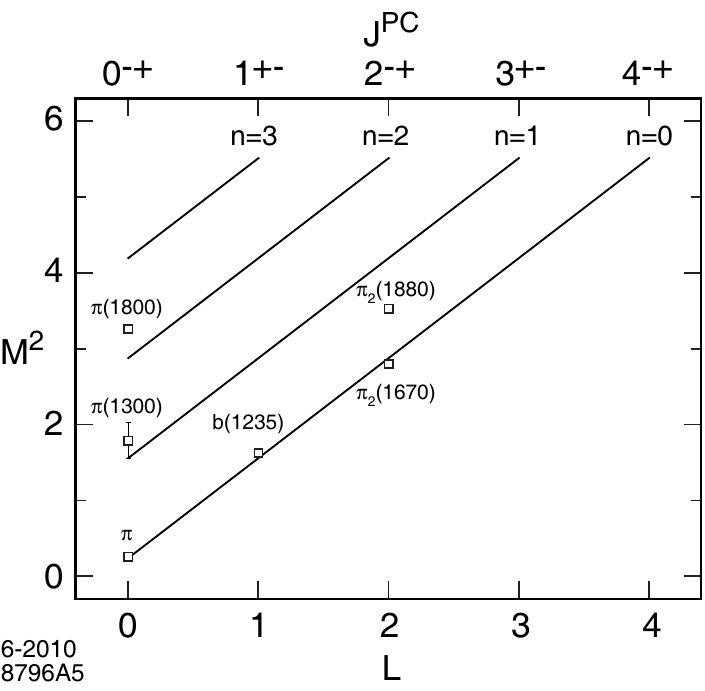}
\caption{Parent and daughter Regge trajectories for the $\pi$-meson family for  
$\kappa= 0.6$ GeV.}
\label{pion}
\end{center}
\end{figure}

\begin{figure}[!]
\begin{center}
\includegraphics[angle=0,width=8.5cm]{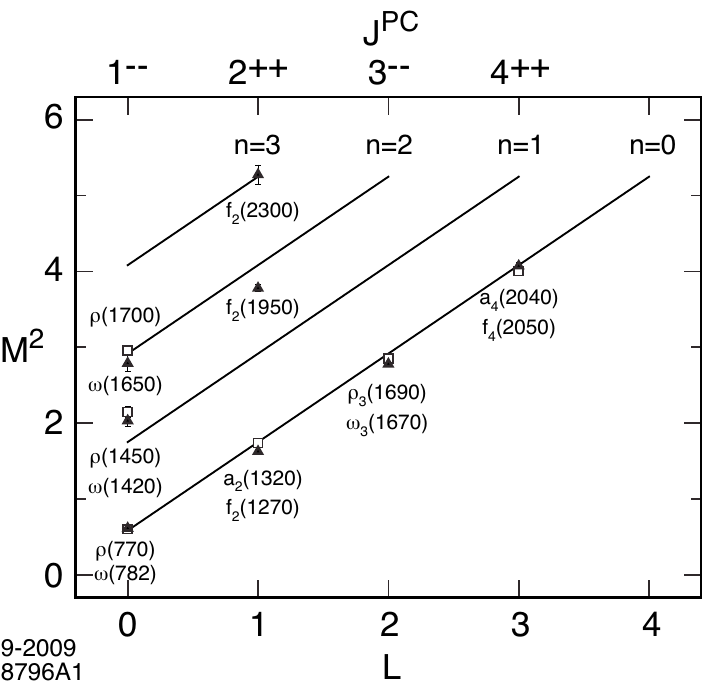}
\caption{Regge trajectories for the  $I\!=\!1$ $\rho$-meson
 and the $I\!=\!0$  $\omega$-meson families for $\kappa= 0.54$ GeV.}
\label{VM}
\end{center}
\end{figure}

The meson spectrum predicted by  Eq. \ref{eq:QCDLFWE} has a string-theory Regge form
${\cal M}^2 = 4 \kappa^2(n+ L+S/2)$; {\it i.e.}, the square of the eigenmasses are linear in both $L$ and $n$, where $n$ counts the number of nodes  of the wavefunction in the radial variable $\zeta$.  This is illustrated for the pseudoscalar and vector meson spectra in Figs. \ref{pion} and \ref{VM},
where the data are from Ref.  \cite{Amsler:2008xx}.
The pion ($S=0, n=0, L=0$) is massless for zero quark mass, consistent with chiral invariance.  Thus one can compute the hadron spectrum by simply adding  $4 \kappa^2$ for a unit change in the radial quantum number, $4 \kappa^2$ for a change in one unit in the orbital quantum number  and $2 \kappa^2$ for a change of one unit of spin $S$. Remarkably, the same rule holds for baryons as we shall show below. For other 
recent calculations of the hadronic spectrum based on AdS/QCD, see Refs.~\cite{BoschiFilho:2005yh,Evans:2006ea, Hong:2006ta, Colangelo:2007pt, Forkel:2007ru, Vega:2008af, Nawa:2008xr,  dePaula:2008fp, Colangelo:2008us, Forkel:2008un, Ahn:2009px, Sui:2009xe, Kapusta:2010mf, Zhang:2010bn, Iatrakis:2010zf, dePaula:2010sy, Branz:2010ub, Kirchbach:2010dm}.

\begin{figure}[!]
\begin{center}
\includegraphics[width=9.5cm]{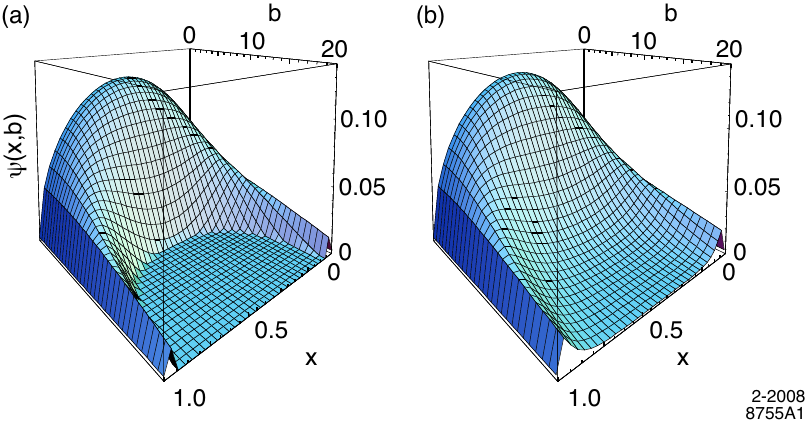}
 \caption{Pion light-front wavefunction $\psi_\pi(x, \mbf{b}_\perp$) for the  AdS/QCD (a) hard-wall ($\Lambda_{QCD} = 0.32$ GeV) and (b) soft-wall  ( $\kappa = 0.375$ GeV)  models.}
\label{LFWF}
\end{center}
\end{figure}

\begin{figure}[h]
\centering
\includegraphics[angle=0,width=8.0cm]{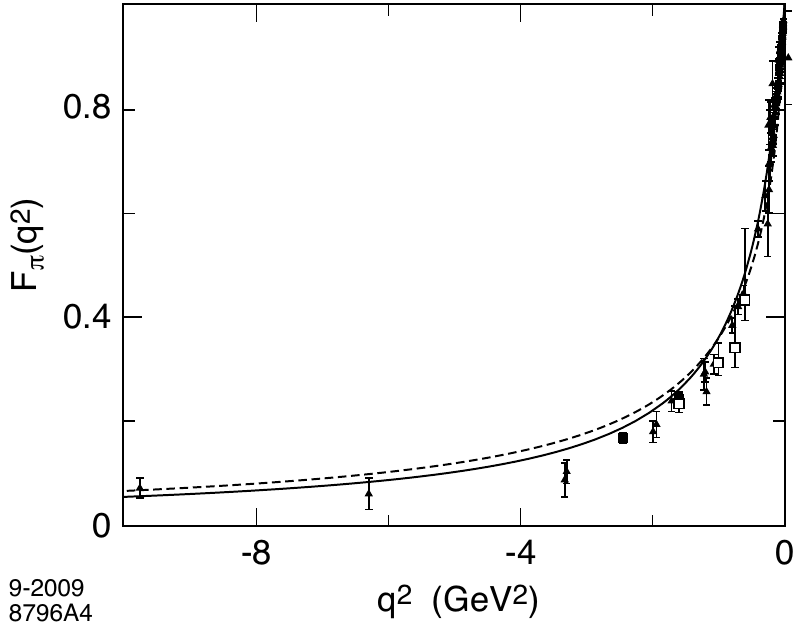}
\caption{Space-like scaling behavior for $F_\pi(Q^2)$ as a function of $q^2$.
The continuous line is the prediction of the soft-wall model for  $\kappa = 0.375$ GeV.
The dashed line is the prediction of the hard-wall model for $\Lambda_{\rm QCD} = 0.22$ GeV.
The triangles are the data compilation  from Baldini  {\it et al.},~\cite{Baldini:1998qn}  the  filled boxes  are JLAB 1 data~\cite{Tadevosyan:2007yd} and empty boxes are JLAB 2
 data.~\cite{Horn:2006tm}}
\label{PionFFSL}
\end{figure}

The eigensolutions of  Eq. \ref{eq:QCDLFWE} provide the light-front wavefunctions of the valence Fock state of the hadrons $
\psi(\zeta) = \psi(x, \mbf{b}_\perp)$
 as illustrated for the pion in Fig. \ref{LFWF} for the soft and hard wall models.  Given these wavefunctions, one can predict many hadronic observables such as the generalized parton distributions that control deeply virtual Compton scattering. For example, hadron form factors can be predicted from the overlap of LFWFs as in the Drell-Yan West formula.
The prediction for the space-like pion form factor is shown in Fig.  \ref{PionFFSL}.

The vector-meson poles residing in the dressed current in the soft-wall model require choosing a value of $\kappa$ smaller by a factor  $1\over \sqrt 2$  than the canonical value of $\kappa$ that determines the mass scale of the hadronic spectra.  This shift is apparently related to the fact that the longitudinal twist-3  current  $J^+$, which appears in the Drell-Yan expression for the space-like form factor,  is different from the  twist-2 $q\bar q$  operator that is used to compute the vector meson spectrum.  The leading-twist operator corresponds to the transverse current $J_\perp$ which is responsible for producing the $\bar q q$ state in the time-region from $e^+  e^-$ annihilation. We will discuss this point further in an upcoming paper.

Other recent computations of the space-like pion form factor in AdS/QCD are presented in~\cite{Kwee:2007dd,Grigoryan:2007wn}. Given
the LFWFs one can compute jet hadronization at the amplitude level from first principles.~\cite{Brodsky:2008tk} A similar method has been used to predict the production of antihydrogen from the off-shell coalescence of relativistic antiprotons and positrons.~\cite{Munger:1993kq}

This semi-classical first approximation to QCD can be systematically improved  by using the complete orthonormal solutions of Eq. \ref{eq:QCDLFWE} to diagonalize the QCD light-front Hamiltonian~\cite{Vary:2009gt}  or by applying the Lippmann-Schwinger method to systematically include the QCD interaction terms. In either case, the result is the full Fock state structure of the hadron eigensolution. One can also model 
heavy-light
and heavy hadrons by including non-zero quark masses in the LF kinetic energy 
$\sum_i ({\mbf{k}^2_{\perp  i}+ m^2_i)/x_i}$
 as well as the effects of the one-gluon exchange potential.

One can derive these results  in two parallel ways. In the first method, one begins with a conformal approximation to QCD, justified by evidence that the theory has an infrared fixed point.~\cite{Deur:2008rf}  One then uses the fact that the conformal group has a geometrical representation in the five-dimensional AdS$_5$ space
to model an effective dual gravity description in AdS. The fact that
conformal invariance is reflected in the isometries of AdS is an essential ingredient of Maldacena's AdS/CFT correspondence.~\cite{Maldacena:1997re} Confinement is
effectively introduced with a sharp cut-off in the infrared region of AdS space, the ``hard-wall" model,~\cite{Polchinski:2001tt}
or with a dilaton background in the fifth dimension which produces a smooth cutoff and linear Regge trajectories,
the ``soft-wall" model.~\cite{Karch:2006pv}
The soft-wall AdS/CFT model with a dilaton-modified AdS space leads to the
 potential $U(z) = \kappa^4 z^2 + 2 \kappa^2(L+S-1)$. This potential can be derived directly from the action in AdS 
 space~\cite{deTeramond:2009xk}
 and corresponds to a dilaton profile $\exp(+\kappa^2 z^2)$, with   a positive argument of the exponential, opposite in sign  to that of Ref.  \cite{Karch:2006pv}.  
 The modified metric induced by the dilaton can be interpreted in AdS space as a gravitational potential
for an object of mass $m$  in the fifth dimension:
$V(z) = mc^2 \sqrt{g_{00}} = mc^2 R \, e^{\pm \kappa^2 z^2/2}/z$.
In the case of the negative solution, the potential decreases monotonically, and thus an object in AdS will fall to infinitely large
values of $z$.  For the positive solution, the potential is non-monotonic and has an absolute minimum at $z_0 = 1/\kappa$.
Furthermore, for large values of $z$ the gravitational potential increases exponentially,  confining any object  to distances
$\langle z \rangle \sim 1/\kappa$~\cite {deTeramond:2009xk}.  We will thus choose the confining positive sign dilaton solution~\cite{deTeramond:2009xk,Andreev:2006ct} with  opposite sign  to that of Ref.~\cite{Karch:2006pv}. This additional warp factor leads to a well-defined scale-dependent effective coupling.

Glazek and Schaden~\cite{Glazek:1987ic} have shown that a  harmonic-oscillator confining potential naturally arises as an effective potential between heavy quark states when one stochastically eliminates higher gluonic Fock states. Also, Hoyer~\cite{Hoyer:2009ep} has argued that the Coulomb  and  linear  potentials are uniquely allowed in the Dirac equation at the lowest  level in $\hbar.$  The linear potential  becomes a harmonic oscillator potential in the corresponding Klein-Gordon equation.

Hadrons are identified in AdS/QCD by matching the power behavior of the hadronic amplitude at the AdS boundary at small $z$ to the twist of its interpolating operator at short distances $x^2 \to 0$, as required by the AdS/CFT dictionary. The twist corresponds to the dimension of fields appearing in chiral super-multiplets.~\cite{Craig:2009rk}
The canonical twist of a hadron equals the number of  its constituents.
We  can then apply light-front holography to relate the amplitude eigensolutions  in the fifth dimension coordinate $z$  to the LF wavefunctions in the physical space-time variable  $\zeta.$  Light-front holography can be derived by establishing an identity between the Polchinski-Strassler formula for current matrix elements in 
AdS space~\cite{Polchinski:2002jw} and the corresponding
Drell-Yan-West formula in LF theory.~\cite{Drell:1969km, West:1970av}  The same correspondence is obtained for both electromagnetic~\cite{Brodsky:2006uqa, Brodsky:2007hb}
 and gravitational form factors,~\cite{Brodsky:2008pf} a nontrivial test of consistency.

In the second method we  use a first semiclassical approximation  to transform the fixed
LF time bound-state Hamiltonian equation  to  a corresponding wave equation in AdS space.
The  invariant LF
coordinate $\zeta$ allows the separation of the dynamics of quark and gluon binding from
the kinematics of constituent spin and internal orbital angular momentum.~\cite{deTeramond:2008ht}  In effect, $\zeta$ represents the off-light-front energy shell ( invariant mass)  dependence of the bound state.
The result is the  single-variable LF relativistic
Schr\"odinger equation  which determines the spectrum
and LFWFs  of hadrons for general spin and
orbital angular momentum. This LF wave equation serves as a semiclassical first
approximation to QCD, and it is equivalent to the
equations of motion which describe the propagation of spin-$J$ modes
in  AdS space.

The term $L^2/ 4 \zeta^2$  in the  LF equation of motion  (\ref{eq:QCDLFWE})
is derived from  the reduction of the LF kinetic energy, when one transforms to the radial $\zeta$  and angular coordinate 
$\varphi$, in analogy to the $\ell(\ell+1)/ r^2$ Casimir term in Schr\"odinger theory.  One thus establishes the interpretation of the orbital angular
momentum $L$ in the AdS equations of motion.
The interaction terms build confinement and correspond to
truncation of AdS space in an effective dual gravity  approximation.~\cite{deTeramond:2008ht}
The duality between these two methods provides a direct
connection between the description of hadronic modes in AdS space and
the Hamiltonian formulation of QCD in physical space-time, quantized
on the light-front  at fixed LF time $\tau.$

\section{Foundations of AdS/QCD}

One of the most significant theoretical advances in recent years has
been the application of the AdS/CFT
correspondence~\cite{Maldacena:1997re} between string theories
defined in 5-dimensional Anti--de Sitter  space-time and
conformal field theories in physical space-time,
to study  the dynamics of strongly coupled quantum field theories.
The essential principle
underlying the AdS/CFT approach to conformal gauge theories is the
isomorphism of the group of Poincar\'{e} and conformal transformations
$SO(4,2)$ to the group of isometries of Anti-de Sitter space.  The
AdS metric is
\begin{equation} \label{eq:AdSz}
ds^2 = \frac{R^2}{z^2}(\eta_{\mu \nu} dx^\mu
 dx^\nu - dz^2),
 \end{equation}
which is invariant under scale changes of the
coordinate in the fifth dimension $z \to \lambda z$ and $ x_\mu \to
\lambda x_\mu$.  Thus one can match scale transformations of the
theory in $3+1$ physical space-time to scale transformations in the
fifth dimension $z$.
In the AdS/CFT duality, the amplitude $\Phi(z)$ represents the
extension of the hadron into the additional fifth dimension.  The
behavior of
$\Phi(z) \to z^\tau$ at $z \to 0$
matches the twist-dimension $\tau$ of the hadron at short distances.

In the standard applications of  AdS/CFT methods, one begins with Maldacena's duality between  the conformal supersymmetric $SO(4,2)$
 gauge theory and a semiclassical supergravity string theory defined in a 10 dimension 
 AdS$_5 \times S^5$
 space-time.  There are no existing
 string theories actually dual to QCD, but nevertheless many interesting predictions can be made. In contrast, in our approach, we simply use the mathematical fact that the effects of scale transformations in a conformal theory can be mapped to the $z$-dependence of amplitudes in AdS$_5$ space.  In this approach, we consider the propagation of hadronic modes in a fixed effective gravitational background which encodes salient properties of the QCD dual theory, such as the ultraviolet conformal limit at the AdS boundary at $z \to 0$, as well as modifications of the background geometry in the large-$z$ infrared region, characteristic of strings dual to confining gauge theories. 
 
The identification of orbital angular momentum of the constituents is a key element in the description of the internal structure of hadrons using holographic principles. In our approach  quark and gluon degrees of freedom are explicitly introduced in the gauge/gravity correspondence, in contrast with the usual
AdS/QCD framework~\cite{Erlich:2005qh,DaRold:2005zs} where axial and vector currents become the primary entities as in effective chiral theory. 
In our approach,
the holographic mapping is carried out in the  strongly coupled regime where QCD is almost conformal corresponding to an infrared fixed-point.
Our analysis follows from developments in light-front
QCD~\cite{deTeramond:2008ht, Brodsky:2006uqa, Brodsky:2007hb, Brodsky:2008pf, Brodsky:2003px, deTeramond:2005su} which have been inspired by the
AdS/CFT correspondence.~\cite{Maldacena:1997re}

QCD is not conformal, but  there is in fact much empirical evidence from lattice gauge theory,~\cite{Furui:2006py} Dyson Schwinger theory,~\cite{vonSmekal:1997is}  and empirical effective charges,~\cite{Deur:2005cf} that the QCD $\beta$-function vanishes in the infrared.~\cite{Deur:2008rf}  The QCD infrared fixed point arises since the propagators of the confined quarks and gluons in the  loop integrals contributing to the $\beta$-function have a maximal wavelength.~\cite{Brodsky:2008be} The decoupling of quantum loops in the infrared is analogous to QED where vacuum polarization corrections to the photon propagator decouple at $Q^2 \to 0$.
Since there is a  window where the QCD coupling is
large and approximately constant,
QCD resembles a conformal theory for massless quarks. Thus, even though QCD is not conformally invariant,
one can use the mathematical representation of the conformal group
in five-dimensional Anti-de Sitter space to construct an analytic
first approximation to the theory.
One then uses AdS$_5$ to represent scale transformations within the conformal window. Unlike the top-down supergravity approach,  one is not limited to hadrons of  spin 
$J \le 2$ in our bottom-up approach, and one can study baryons with $N_C=3.$
The theory also predicts
dimensional counting for form factors and other fixed CM angle exclusive reactions.  Moreover, as we shall review,
light-front holography allows one to map the hadronic amplitudes $\Phi(z)$ determined in the AdS fifth dimension $z$ to the valence LFWFs of each hadron as a function of a covariant impact variable $\zeta.$  Moreover, the same techniques provide a prediction for the QCD coupling $\alpha_s(Q^2)$ and its $\beta$-function which reflects the dynamics of confinement.

\section{Light-Front Quantization}

Light-front quantization is the ideal framework for  describing the
structure of hadrons in terms of their quark and gluon degrees of
freedom. The light-front wavefunctions
of bound states in QCD are relativistic generalizations of the Schr\"odinger wavefunctions, but they are determined at fixed light-front time $\tau = x^+ = x^0 + x^3$, the time marked by the
front of a light wave~\cite{Dirac:1949cp}, rather than at fixed ordinary time
$t.$ They play the same role in
hadron physics that Schr\"odinger wavefunctions play in atomic physics.
In addition, the simple structure of the LF vacuum provides an unambiguous
definition of the partonic content of a hadron in QCD.

We can  define the  LF Lorentz invariant Hamiltonian
$H_{LF}= P_\mu P^\mu = P^-P^+  \! - \mbf{P}^2_\perp$ with eigenstates
$\vert \psi_H(P^+, \mbf{P}_\perp, S_z )\rangle$
and eigenmass  $\mathcal{M}_H^2$, the mass spectrum of the color-singlet states
of QCD.~\cite{Brodsky:1997de}
$H_{LF}$ can be determined canonically from the QCD Lagrangian in light-cone gauge $A^+=0$.  The light-front formalism for gauge theories in
light-cone gauge is particularly useful in that there are no ghosts
and one has a direct physical interpretation of  orbital angular
momentum. The Heisenberg equation for QCD on the light-front thus takes the form $H_{LF} \vert \psi_H \rangle = \mathcal{M}^2_H \vert \psi_H \rangle $.
Its eigenfunctions are the light-front eigenstates which define the frame-independent light-front wavefunctions, and
its eigenvalues yield the hadronic spectrum, the bound states as well as the continuum.  
A state $\vert \psi_H \rangle$ is an expansion
in multi-particle Fock states
$\vert n \rangle $ of the free LF Hamiltonian:
~$\vert \psi_H \rangle = \sum_n \psi_{n/H} \vert n \rangle$, where
a one parton state is $\vert q \rangle = \sqrt{2 q^+} \,b^\dagger(q) \vert 0 \rangle$.
The  projection of the eigensolutions on the free Fock basis thus give the $n$-parton LF wavefunctions
$\psi_{n/H} = \langle n \vert \psi_H \rangle$ needed for phenomenology.

Light-front quantization of QCD provides a nonperturbative method for  solving  QCD in Minkowski space. Unlike lattice gauge theories, fermions introduce no new complications.
Heisenberg's problem on the light-front can be solved numerically using discretized light-front quantization (DLCQ)~\cite{Pauli:1985ps} by applying anti-periodic boundary conditions in $\sigma = x^0-x^3.$  This method has been used successfully to solve many lower dimension quantum field theories.~\cite{Brodsky:1997de}

\section{Light-Front Wavefunctions}

The Schr\"odinger wavefunction describes the quantum-mechanical
structure of  an atomic system at the amplitude level. The light-front multiparticle Fock state
wavefunctions play a similar role in quantum chromodynamics,
providing a fundamental description of the structure and
internal dynamics of hadrons in terms of their constituent quarks
and gluons. The LFWFs of bound states in QCD are
relativistic generalizations of the Schr\"odinger wavefunctions of
atomic physics, but they are determined at fixed light-cone time
$\tau  = x^0 +x^3/c$ -- the ``front form'' introduced by
Dirac~\cite{Dirac:1949cp} -- rather than at fixed ordinary time $t.$

When a flash from a camera illuminates a scene, each object is illuminated along the light-front of the flash; i.e., at a given $\tau$.  Similarly, when a sample is illuminated by an x-ray source, each element of the target is struck at a given $\tau.$  In contrast, setting the initial condition using conventional instant time $t$ requires simultaneous scattering of photons on each constituent.
Thus it is natural to set boundary conditions at fixed $\tau$ and then evolve the system using the light-front  Hamiltonian
$P^-  \!= \!  P^0-P^3 = i {d/d \tau}.$  The invariant Hamiltonian $H_{LF} = P^+ P^- \! - \mbf{P}^2_\perp$ then has eigenvalues $\mathcal{M}^2$ where $\mathcal{M}$ is the physical mass.   Its eigenfunctions are the light-front eigenstates whose Fock state projections define the light-front wavefunctions.  Given the LF Fock state wavefunctions
$\psi^H_n(x_i, \mbf{k}_{\perp i}, \lambda_i),$
where $x_i \! =\! k^+/P^+$,  $\sum_{i=1}^n x_i  \! = \! 1, ~ \sum_{i=1}^n \mbf{k}_{\perp i}  \! = \! 0$, one
can immediately compute observables such as hadronic form factors (overlaps of LFWFs), structure
functions (squares of LFWFS), as well as the generalized parton distributions and
distribution amplitudes which underly hard exclusive reactions.

The most useful  feature of LFWFs is the fact that they are frame-independent; i.e., the form of the LFWF is independent of the
hadron's total momentum $P^+ = P^0 + P^z$ and $\mbf{P}_\perp.$
The simplicity of Lorentz boosts of LFWFs contrasts dramatically with the complexity of the boost of wavefunctions defined at fixed time $t.$~\cite{Brodsky:1968ea}
Light-front quantization is thus the ideal framework to describe the
structure of hadrons in terms of their quark and gluon degrees of freedom.  The
constituent spin and orbital angular momentum properties of the
hadrons are also encoded in the LFWFs.
The total  angular momentum projection,~\cite{Brodsky:2000ii}
$J^z = \sum_{i=1}^n  S^z_i + \sum_{i=1}^{n-1} L^z_i$,
is conserved Fock-state by Fock-state and by every interaction in the LF Hamiltonian.

The angular momentum projections in the LF $\hat z$ direction $L^z, S^z$ and $J^z$ are kinematical in the front form, so they are the natural quantum numbers to label the eigenstates of light-front  physics.  In the massless fermion limit, the gauge interactions conserve LF chirality;  i.e, the quark spin $S^z$ is conserved (rather than helicity $\vec S \cdot \vec p$) at the QCD vertices. Similarly, quark-antiquark pairs from gluon splitting have opposite $S^z.$    For example,  the light-front  wavefunction of a massless electron at order $\alpha$ in QED with $J^z= +1/2$ has fermion-photon Fock components $S^z_e = +1/2, S^z_\gamma=+1, L^z=-1$  and $S^z_e = +1/2, S^z_\gamma=-1, L^z=+1.$

In general, a hadronic eigenstate with spin $ j$ in the front form corresponds to an eigenstate of $J^2 = j(j+1)$ in the rest frame in the conventional instant form.
It thus has   $2 j + 1 $ degenerate states with $J^z= - j, -j+1, \cdots j-1, +j$.~\cite{Brodsky:1997de}  An important feature of relativistic theories is that hadron eigenstates have in general Fock components with different $L$ components. For example, the  $S^z = +1/2$  proton eigenstate in 
AdS space has equal probability 
$S^z_q = +1/2, L^z=0$ and $S^z_q= -1/2, L^z=+1$ light-front Fock components with equal probability.
Thus in AdS, the proton is effectively a quark/scalar-diquark composite with equal $S$ and P waves.

In the case of QED, the ground state $1S$ state of the Dirac-Coulomb equation has both $L=0$ and $L=1$ components.   By convention, in both light-front QCD and QED, one labels the eigenstate with its minimum value of $L$.  For example, the symbol $L$ in the AdS/QCD spectral  prediction $M^2 = 4 \kappa^2(n + L + S/2)$ refers to the {\it minimum } $L$, and $S= j$ is the total spin of the hadron.

Other advantageous features of light-front quantization include:

\begin{itemize}

\item
The simple structure of the light-front vacuum allows an unambiguous
definition of the partonic content of a hadron in QCD.  The chiral and gluonic condensates are properties of the higher Fock states,~\cite{Casher:1974xd,Brodsky:2009zd} rather than the vacuum.  In the case of the Higgs model, the effect of the usual Higgs vacuum expectation value is replaced by a constant $k^+=0$ zero mode field.~\cite{Srivastava:2002mw}  The LF vacuum is trivial up to zero modes in the front form, thus eliminating contributions to the cosmological constant from QED or QCD.~\cite{Brodsky:2009zd} We discuss the remarkable consequences of this for the cosmological constant in section 14.

\item
If one quantizes QCD in the physical light-cone gauge (LCG) $A^+ =0$, then gluons only have physical angular momentum projections $S^z= \pm 1$. The orbital angular momenta of quarks and gluons are defined unambiguously, and there are no ghosts.

\item
The gauge-invariant distribution amplitude $\phi(x,Q)$  is the integral of the valence LFWF in LCG integrated over the internal transverse momentum $k^2_\perp < Q^2$ because the Wilson line is trivial in this gauge. It is also possible to quantize QCD in  Feynman gauge in the light front.~\cite{Srivastava:1999gi}

\item
LF Hamiltonian perturbation theory provides a simple method for deriving analytic forms for the analog of Parke-Taylor amplitudes~\cite{Motyka:2009gi} where each particle spin $S^z$ is quantized in the LF $z$ direction.  The gluonic $g^6$ amplitude  $T(-1 -1 \to +1 +1 +1 +1 +1 +1)$  requires $\Delta L^z =8;$ it thus must vanish at tree level since each three-gluon vertex has  $\Delta L^z = \pm 1.$ However, the order $g^8$ one-loop amplitude can be nonzero.

\item
Amplitudes in light-front perturbation theory may be automatically renormalized using the ``alternate denominator"  subtraction method.~\cite{Brodsky:1973kb} The application to QED has been checked at one and two loops.~\cite{Brodsky:1973kb}

\item
A fundamental theorem for gravity can be derived from the equivalence principle:  the anomalous gravitomagnetic moment defined from the spin-flip  matrix element of the energy-momentum tensor is identically zero, $B(0)=0$.~\cite{Teryaev:1999su} This theorem can be proven in  the light-front formalism Fock state by Fock state.~\cite{Brodsky:2000ii} 

\item
LFWFs obey the cluster decomposition theorem, providing an elegant proof of this theorem for relativistic bound states.~\cite{Brodsky:1985gs}

\item
The LF Hamiltonian can be diagonalized using the discretized light-cone quantization (DLCQ) method.~\cite{Pauli:1985ps} This nonperturbative method is particularly useful for solving low-dimension quantum field theories such as
QCD$(1+1).$~\cite{Hornbostel:1988fb}

\item
LF quantization provides a distinction between static  (the square of LFWFs) distributions versus non-universal dynamic structure functions,  such as the Sivers single-spin correlation and diffractive deep inelastic scattering which involve final state interactions.  The origin of nuclear shadowing and process independent anti-shadowing also becomes explicit.

\item
LF quantization provides a simple method to implement jet hadronization at the amplitude level.

\item
The instantaneous fermion interaction in LF  quantization provides a simple derivation of the $J=0$
fixed pole contribution to deeply virtual Compton scattering,~\cite{Brodsky:2009bp} i.e., the $e^2_q s^0 F(t)$  contribution to the DVCS amplitude which is independent of photon energy and virtuality.

\item
Unlike instant time quantization, the bound state
Hamiltonian equation of motion in the LF is frame independent. This makes a direct connection of QCD with AdS/CFT methods possible.~\cite{deTeramond:2008ht}

\end{itemize}

\section{Applications of Light-Front Wavefunctions}

The Fock components $\psi_{n/H}(x_i, {\mathbf{k}_{\perp i}}, \lambda_i^z)$
are independent of  $P^+$ and $\mbf{P}_{\! \perp}$
and depend only on relative partonic coordinates:
the momentum fraction
 $x_i = k^+_i/P^+$, the transverse momentum  ${\mathbf{k}_{\perp i}}$ and spin
 component $\lambda_i^z$.
 The LFWFs $\psi_{n/H}$ provide a
{\it frame-independent } representation of a hadron which relates its quark
and gluon degrees of freedom to their asymptotic hadronic state.  Since the LFWFs are independent of the hadron's total momentum $P^+ = P^0 \! + P^3$, so that once they are known in one frame, they are known in all frames; Wigner transformations and Melosh rotations
are not required.   They also allow one to formulate hadronization in inclusive and exclusive reactions at the amplitude level.

A key example of the utility of the light-front formalism is the Drell-Yan West formula~\cite{Drell:1969km, West:1970av}
for the space like form factors of electromagnetic currents given as overlaps of initial and final LFWFs. At high momentum where one can iterate the hard scattering kernel, this yields the dimensional counting rules, factorization theorems, and ERBL evolution of the distribution amplitudes. The gauge-invariant distribution
amplitudes $\phi_H(x_i,Q)$ defined from the integral over the
transverse momenta $\mbf{k}^2_{\perp i} \le Q^2$ of the valence
(smallest $n$) Fock state provide a fundamental measure of the
hadron at the amplitude level;~\cite{Lepage:1979zb, Efremov:1979qk}
they  are the nonperturbative inputs to the factorized form of hard
exclusive amplitudes and exclusive heavy hadron decays in
pQCD.

Given the light-front wavefunctions $\psi_{n/H}$ one can
compute a large range of other hadron
observables. For example, the valence and sea quark and gluon
distributions which are measured in deep inelastic lepton scattering
are defined from the squares of the LFWFs summed over all Fock
states $n$. Exclusive weak transition
amplitudes~\cite{Brodsky:1998hn} such as $B\to \ell \nu \pi$,  and
the generalized parton distributions~\cite{Brodsky:2000xy} measured
in deeply virtual Compton scattering  $\gamma^* p \to \gamma p$ are (assuming the ``handbag"
approximation) overlap of the initial and final LFWFs with $n
=n^\prime$ and $n =n^\prime+2$. The resulting distributions obey the DGLAP and
ERBL evolution equations as a function of the maximal invariant
mass, thus providing a physical factorization
scheme.~\cite{Lepage:1980fj} In each case, the derived quantities
satisfy the appropriate operator product expansions, sum rules, and
evolution equations. At large $x$ where the struck quark is
far-off shell, DGLAP evolution is quenched,~\cite{Brodsky:1979qm} so
that the fall-off of the DIS cross sections in $Q^2$ satisfies Bloom-Gilman
inclusive-exclusive duality at fixed $W^2.$

The simple features of the light-front contrast with the conventional instant form where one quantizes at $t=0.$ For example, calculating a hadronic form factor requires boosting the hadron's wavefunction from the initial to final state, a dynamical problem  as difficult as solving QCD itself. Moreover current matrix elements require computing the interaction of the probe with all of connected currents fluctuating in the QCD vacuum.  Each contributing diagram is frame-dependent.

\section{Light-Front Holography}

Light-front  holography \cite{deTeramond:2008ht, Brodsky:2006uqa, Brodsky:2007hb, Brodsky:2008pf,  deTeramond:2010we} 
 maps a confining gauge theory quantized on the light front to a higher-dimensional anti-de Sitter space  incorporating the AdS/CFT correspondence as a useful guide.
 This correspondence provides a direct connection between the hadronic amplitudes $\Phi(z)$  in AdS space  with  LF wavefunctions $\phi(\zeta)$ describing the quark and gluon constituent structure of hadrons in physical space-time.
In the case of a meson, $\zeta = \sqrt{x(1-x) {\bf b}^2_\perp}$ is a Lorentz invariant coordinate which measures
the distance between the quark and antiquark; it is analogous to the radial coordinate $r$ in the Schr\"odinger equation. Here $\mbf{b}_\perp$ is the Fourier conjugate of the transverse
momentum $\mbf{k}_\perp$. The variable $\zeta$ also  represents the off-light-front energy shell and invariant mass dependence of the bound state. Light-front holography provides a  connection between the description of
hadronic modes in AdS space and the Hamiltonian formulation of QCD in
physical space-time quantized on the light-front  at fixed LF
time $\tau.$ The resulting equation for the mesonic $q \bar q$ bound states at fixed light-front time, Eq. (\ref{eq:QCDLFWE}),  has the form of a single-variable relativistic Lorentz invariant Schr\"odinger equation.~\cite{deTeramond:2008ht} 
The resulting light-front eigenfunctions provide a fundamental description of the structure and internal dynamics of hadronic states in terms of their constituent quark and gluons.

\section{Holographic Mapping of Transition Amplitudes}

The mapping between the LF invariant variable $\zeta$ and the fifth-dimension AdS coordinate $z$ was originally obtained
by matching the expression for electromagnetic current matrix
elements in AdS space  with the corresponding expression for the
current matrix element, using LF  theory in physical space
time.~\cite{Brodsky:2006uqa} It has also been shown that one
obtains the identical holographic mapping using the matrix elements
of the energy-momentum tensor,~\cite{Brodsky:2008pf,Abidin:2008ku} thus  verifying  the  consistency of the holographic
mapping from AdS to physical observables defined on the light front.

The light-front electromagnetic form factor in impact
space~\cite{Brodsky:2006uqa,Brodsky:2007hb,Soper:1976jc} can be written as a sum of overlap of light-front wave functions of the $j = 1,2, \cdots, n-1$ spectator
constituents:
\begin{equation} \label{eq:FFb}
F(q^2) =  \sum_n  \prod_{j=1}^{n-1}\int d x_j d^2 \mbf{b}_{\perp j}   \sum_q e_q
            \exp \! {\Bigl(i \mbf{q}_\perp \! \cdot \sum_{j=1}^{n-1} x_j \mbf{b}_{\perp j}\Bigr)}
 \left\vert  \psi_{n/H}(x_j, \mbf{b}_{\perp j})\right\vert^2
\end{equation}
where the normalization is defined by
\vspace{-4pt}
\begin{equation}  \label{eq:Normb}
\sum_n  \prod_{j=1}^{n-1} \int d x_j d^2 \mathbf{b}_{\perp j}
\vert \psi_{n/H}(x_j, \mathbf{b}_{\perp j})\vert^2 = 1.
\end{equation}

The formula  is exact if the sum is over all Fock states $n$.
For definiteness we shall consider a two-quark $\pi^+$  valence Fock state
$\vert u \bar d\rangle$ with charges $e_u = \frac{2}{3}$ and $e_{\bar d} = \frac{1}{3}$.
For $n=2$, there are two terms which contribute to the $q$-sum in (\ref{eq:FFb}).
Exchanging $x \leftrightarrow 1 \! - \! x$ in the second integral  we find
\begin{equation}  \label{eq:PiFFb}
 F_{\pi^+}(q^2)  =  2 \pi \int_0^1 \! \frac{dx}{x(1-x)}  \int \zeta d \zeta
J_0 \! \left(\! \zeta q \sqrt{\frac{1-x}{x}}\right)
\left\vert \psi_{u \bar d/ \pi}\!(x,\zeta)\right\vert^2,
\end{equation}
where $\zeta^2 =  x(1  -  x) \mathbf{b}_\perp^2$ and $F_{\pi^+}(q\!=\!0)=1$.

We now compare this result with the electromagnetic form-factor in AdS space:~\cite{Polchinski:2002jw}
\begin{equation}
F(Q^2) = R^3 \int \frac{dz}{z^3} \, J(Q^2, z) \vert \Phi(z) \vert^2,
\label{eq:FFAdS}
\end{equation}
where $J(Q^2, z) = z Q K_1(z Q)$.
Using the integral representation of $J(Q^2,z)$
\begin{equation} \label{eq:intJ}
J(Q^2, z) = \int_0^1 \! dx \, J_0\negthinspace \left(\negthinspace\zeta Q
\sqrt{\frac{1-x}{x}}\right) ,
\end{equation} we write the AdS electromagnetic form-factor as
\begin{equation}
F(Q^2)  =    R^3 \! \int_0^1 \! dx  \! \int \frac{dz}{z^3} \,
J_0\!\left(\!z Q\sqrt{\frac{1-x}{x}}\right) \left \vert\Phi(z) \right\vert^2 .
\label{eq:AdSFx}
\end{equation}
Comparing with the light-front QCD  form factor (\ref{eq:PiFFb}) for arbitrary  values of $Q$~\cite{Brodsky:2006uqa}
\begin{equation} \label{eq:Phipsi}
\vert \psi(x,\zeta)\vert^2 =
\frac{R^3}{2 \pi} \, x(1-x)
\frac{\vert \Phi(\zeta)\vert^2}{\zeta^4},
\end{equation}
where we identify the transverse LF variable $\zeta$, $0 \leq \zeta \leq \Lambda_{\rm QCD}$,
with the holographic variable $z$.
Identical results are obtained from the mapping of the QCD gravitational form factor
with the expression for the hadronic gravitational form factor in AdS space.~\cite{Brodsky:2008pf,Abidin:2008ku}

\section{A Semiclassical Approximation to QCD}

One can also derive light-front holography using a first semiclassical approximation  to transform the fixed
light-front time bound-state Hamiltonian equation of motion in QCD  to  a corresponding wave equation in AdS
space.~\cite{deTeramond:2008ht} To this end we
 compute the invariant hadronic mass $\mathcal{M}^2$ from the hadronic matrix element
\begin{equation}
\langle \psi_H(P') \vert H_{LF}\vert\psi_H(P) \rangle  =
\mathcal{M}_H^2  \langle \psi_H(P' ) \vert\psi_H(P) \rangle,
\end{equation}
expanding the initial and final hadronic states in terms of its Fock components. We use the
frame $P = \big(P^+, M^2/P^+, \vec{0}_\perp \big)$ where $H_{LF} =  P^+ P^-$.
The LF expression for $\mathcal{M}^2$ In impact space is
\begin{multline}
 \mathcal{M}_H^2  =  \sum_n  \prod_{j=1}^{n-1} \int d x_j \, d^2 \mbf{b}_{\perp j} \,
\psi_{n}^*(x_j, \mbf{b}_{\perp j})    \,
 \sum_q   \left(\frac{ \mbf{- \nabla}_{ \mbf{b}_{\perp q}}^2  \! + m_q^2 }{x_q} \right)
 \psi_{n}(x_j, \mbf{b}_{\perp j}) \\
  + {\rm (interactions)} , \label{eq:Mb}
 \end{multline}
plus similar terms for antiquarks and gluons ($m_g = 0)$.

To simplify the discussion we will consider a two-parton hadronic bound state.  In the limit
of zero quark mass
$m_q \to 0$
\begin{equation}  \label{eq:MKb}
\mathcal{M}^2  =  \int_0^1 \! \frac{d x}{x(1-x)} \int  \! d^2 \mbf{b}_\perp  \,
  \psi^*(x, \mbf{b}_\perp)
  \left( - \mbf{\nabla}_{ {\mbf{b}}_{\perp}}^2\right)
  \psi(x, \mbf{b}_\perp) +   {\rm (interactions)}.
 \end{equation}
 The functional dependence  for a given Fock state is
given in terms of the invariant mass
\begin{equation}
 \mathcal{M}_n^2  = \Big( \sum_{a=1}^n k_a^\mu\Big)^2 = \sum_a \frac{\mbf{k}_{\perp a}^2 +  m_a^2}{x_a}
 \to \frac{\mbf{k}_\perp^2}{x(1-x)} \,,
 \end{equation}
 the measure of the off-mass shell energy~ $\mathcal{M}^2 - \mathcal{M}_n^2$
 of the bound state.
 Similarly in impact space the relevant variable for a two-parton state is  $\zeta^2= x(1-x)\mbf{b}_\perp^2$.
Thus, to first approximation  LF dynamics  depend only on the boost invariant variable
$\mathcal{M}_n$ or $\zeta,$
and hadronic properties are encoded in the hadronic mode $\phi(\zeta)$ from the relation
\begin{equation} \label{eq:psiphi}
\psi(x,\zeta, \varphi) = e^{i M \varphi} X(x) \frac{\phi(\zeta)}{\sqrt{2 \pi \zeta}} ,
\end{equation}
thus factoring out the angular dependence $\varphi$ and the longitudinal, $X(x)$, and transverse mode $\phi(\zeta)$
with normalization $ \langle\phi\vert\phi\rangle = \int \! d \zeta \,
 \vert \langle \zeta \vert \phi\rangle\vert^2 = 1$.

We can write the Laplacian operator in (\ref{eq:MKb}) in circular cylindrical coordinates $(\zeta, \varphi)$
and factor out the angular dependence of the
modes in terms of the $SO(2)$ Casimir representation $L^2$, $L = L^z$, of orbital angular momentum in the
transverse plane. Using  (\ref{eq:psiphi}) we find~\cite{deTeramond:2008ht}
\begin{equation} \label{eq:KV}
\mathcal{M}^2   =  \int \! d\zeta \, \phi^*(\zeta) \sqrt{\zeta}
\left( -\frac{d^2}{d\zeta^2} -\frac{1}{\zeta} \frac{d}{d\zeta}
+ \frac{L^2}{\zeta^2}\right)
\frac{\phi(\zeta)}{\sqrt{\zeta}}
+ \int \! d\zeta \, \phi^*(\zeta) U(\zeta) \phi(\zeta) ,
\end{equation}
where all the complexity of the interaction terms in the QCD Lagrangian is summed up in the effective potential $U(\zeta)$.
The LF eigenvalue equation $H_{LF} \vert \phi \rangle  =  \mathcal{M}^2 \vert \phi \rangle$
is thus a light-front  wave equation for $\phi$
\begin{equation} \label{eq:QCDLFWE2}
\left(-\frac{d^2}{d\zeta^2}
- \frac{1 - 4L^2}{4\zeta^2} + U(\zeta) \right)
\phi(\zeta) = \mathcal{M}^2 \phi(\zeta),
\end{equation}
an effective single-variable light-front Schr\"odinger equation which is
relativistic, covariant and analytically tractable. 
It is important to notice that in the light-front  the $SO(2)$ Casimir for orbital angular momentum $L^2$
is a kinematical quantity, in contrast with the usual $SO(3)$ Casimir $\ell(\ell+1)$ from non-relativistic physics which is
rotational, but not boost invariant. Using (\ref{eq:Mb}) one can readily
generalize the equations to allow for the kinetic energy of massive
quarks.~\cite{Brodsky:2008pg}  In this case, however,
the longitudinal mode $X(x)$ does not decouple from the effective LF bound-state equations.

As the simplest example, we consider a bag-like model
where  partons are free inside the hadron
and the interaction terms effectively build confinement. The effective potential is a hard wall:
$U(\zeta) = 0$ if  $\zeta \le 1/\Lambda_{\rm QCD}$ and
 $U(\zeta) = \infty$ if $\zeta > 1/\Lambda_{\rm QCD}$,
 where boundary conditions are imposed on the
 boost invariant variable $\zeta$ at fixed light-front time.   If $L^2 \ge 0$ the LF Hamiltonian is positive definite
 $\langle \phi \vert H_{LF} \vert \phi \rangle \ge 0$ and thus $\mathcal M^2 \ge 0$.
 If $L^2 < 0$ the bound state equation is unbounded from below and the particle
 ``falls towards the center''. The critical value corresponds to $L=0$.
  The mode spectrum  follows from the boundary conditions
 $\phi \! \left(\zeta = 1/\Lambda_{\rm QCD}\right) = 0$, and is given in
 terms of the roots of Bessel functions: $\mathcal{M}_{L,k} = \beta_{L, k} \Lambda_{\rm QCD}$.
 Upon the substitution $\Phi(\zeta) \sim \zeta^{3/2} \phi(\zeta)$, $\zeta \to z$
 we find
 \begin{equation} \label{eq:eomPhiJz}
\left[ z^2 \partial_z^2 - 3 z \, \partial_z + z^2 \mathcal{M}^2
\!  -  (\mu R)^2 \right] \!  \Phi_J  = 0,
\end{equation}
 the wave equation which describes the propagation of a scalar mode in a fixed AdS$_5$ background with AdS radius $R$.
 The five dimensional mass $\mu$ is related to the orbital angular momentum of the hadronic bound state by
 $(\mu R)^2 = - 4 + L^2$. The quantum mechanical stability $L^2 >0$ is thus equivalent to the
 Breitenlohner-Freedman stability bound in AdS.~\cite{Breitenlohner:1982jf}
The scaling dimensions are $\Delta = 2 + L$ independent of $J$ in agreement with the
twist scaling dimension of a two parton bound state in QCD. Higher spin-$J$ wave equations
are obtained by shifting dimensions: $\Phi_J(z) = (z/R)^{-J} \Phi(z)$.~\cite{deTeramond:2008ht}

The hard-wall LF model discussed here is equivalent to the hard wall model of
 Ref.~\cite{Polchinski:2001tt}.   The variable $\zeta$,
 $0 \leq \zeta \leq \Lambda_{\rm QCD}^{-1}$,  represents the invariable separation between pointlike constituents and is also
 the holographic variable $z$ in AdS, thus we can identify $\zeta = z$.
 Likewise a two-dimensional  oscillator with
 effective potential  $ U(z) = \kappa^4 z^2 + 2 \kappa^2(L+S-1)$ is similar to the soft-wall model of
 Ref.~\cite{Karch:2006pv} which reproduce the usual linear Regge trajectories, where $L$ is the internal
 orbital angular momentum and $S$ is the internal spin. The soft-wall discussed here correspond to a positive sign
 dilaton and higher-spin solutions also follow from shifting 
 dimensions: $\Phi_J(z) = (z/R)^{-J} \Phi(z)$.~\cite{deTeramond:2009xk}

Individual hadron states can be identified by their interpolating operator at $z\to 0.$  For example, the pseudoscalar meson interpolating operator
$\mathcal{O}_{2+L}= \bar q \gamma_5 D_{\{\ell_1} \cdots D_{\ell_m\}} q$,
written in terms of the symmetrized product of covariant
derivatives $D$ with total internal  orbital
momentum $L = \sum_{i=1}^m \ell_i$, is a twist-two, dimension $3 + L$ operator
with scaling behavior determined by its twist-dimension $ 2 + L$. Likewise
the vector-meson operator
$\mathcal{O}_{2+L}^\mu = \bar q \gamma^\mu D_{\{\ell_1} \cdots D_{\ell_m\}} q$
has scaling dimension $\Delta=2 + L$.  The scaling behavior of the scalar and vector AdS modes $\Phi(z) \sim z^\Delta$ at $z \to 0$  is precisely the scaling required to match the scaling dimension of the local pseudoscalar and vector-meson interpolating operators.
The spectral predictions for the light pseudoscalar and
vector mesons  in the Chew-Frautschi plot in Fig. \ref{pion} and \ref{VM} for the soft-wall model discussed here are in good agreement for the principal and daughter Regge trajectories. Radial excitations correspond to the nodes
of the wavefunction.

\section{Baryons in Light-Front Holography}

For baryons, the light-front wave equation is a linear equation
determined by the LF transformation properties of spin 1/2 states. A linear confining potential
$U(\zeta) \sim \kappa^2 \zeta$ in the LF Dirac
equation leads to linear Regge trajectories.~\cite{Brodsky:2008pg}   For fermionic modes the light-front matrix
Hamiltonian eigenvalue equation $D_{LF} \vert \psi \rangle = \mathcal{M} \vert \psi \rangle$, $H_{LF} = D_{LF}^2$,
in a $2 \times 2$ spinor  component
representation is equivalent to the system of coupled linear equations
\begin{eqnarray} \label{eq:LFDirac} \nonumber
- \frac{d}{d\zeta} \psi_- -\frac{\nu+\half}{\zeta}\psi_-
- \kappa^2 \zeta \psi_-&=&
\mathcal{M} \psi_+, \\ \label{eq:cD2k}
  \frac{d}{d\zeta} \psi_+ -\frac{\nu+\half}{\zeta}\psi_+
- \kappa^2 \zeta \psi_+ &=&
\mathcal{M} \psi_-.
\end{eqnarray}
with eigenfunctions
\begin{eqnarray} \nonumber
\psi_+(\zeta) &\sim& z^{\frac{1}{2} + \nu} e^{-\kappa^2 \zeta^2/2}
  L_n^\nu(\kappa^2 \zeta^2) ,\\
\psi_-(\zeta) &\sim&  z^{\frac{3}{2} + \nu} e^{-\kappa^2 \zeta^2/2}
 L_n^{\nu+1}(\kappa^2 \zeta^2),
\end{eqnarray}
and  eigenvalues
\begin{equation}
\mathcal{M}^2 = 4 \kappa^2 (n + \nu + 1) .
\end{equation}

The baryon interpolating operator
$ \mathcal{O}_{3 + L} =  \psi D_{\{\ell_1} \dots
 D_{\ell_q } \psi D_{\ell_{q+1}} \dots
 D_{\ell_m\}} \psi$,  $L = \sum_{i=1}^m \ell_i$ is a twist 3,  dimension $9/2 + L$ with scaling behavior given by its
 twist-dimension $3 + L$. We thus require $\nu = L+1$ to match the short distance scaling behavior. Higher spin fermionic modes are obtained by shifting dimensions for the fields as in the bosonic case.
Thus, as in the meson sector,  the increase  in the 
mass squared for  higher baryonic state is
$\Delta n = 4 \kappa^2$, $\Delta L = 4 \kappa^2$ and $\Delta S = 2 \kappa^2,$ 
relative to the lowest ground state,  the proton.
Since our starting point to find the bound state equation of motion for baryons is the light-front, we fix the overall energy scale identical for mesons and baryons by imposing chiral symmetry to the pion~\cite{deTeramond:2010we} in the LF Hamiltonian equations. By contrast, if we start with a five-dimensional action for a scalar field in presence of a positive sign dilaton, the pion is automatically massless.

\begin{figure}[!]
\begin{center}
\includegraphics[angle=0,width=12.5cm]{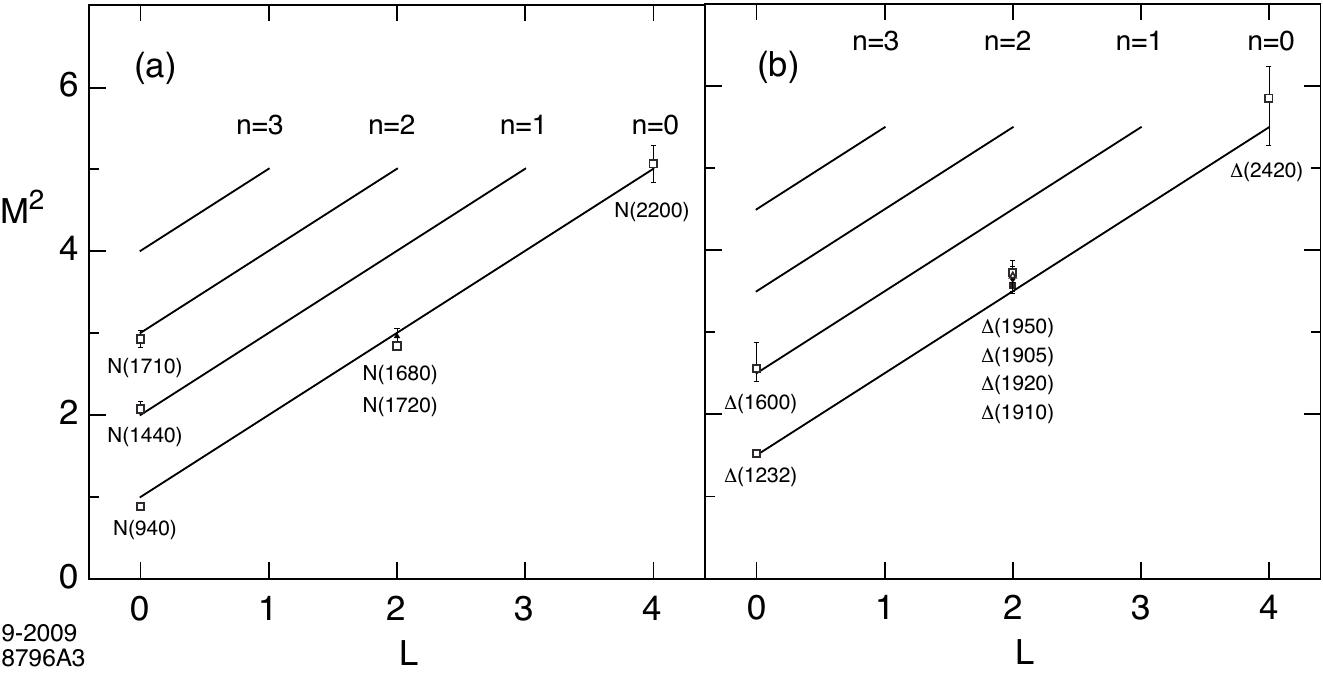}~~~
\caption{{{\bf 56} Regge trajectories for  the  $N$ and $\Delta$ baryon families for $\kappa= 0.5$ GeV}
}
\label{Baryons}
\end{center}
\end{figure}

The predictions for the $\bf 56$-plet of light baryons under the $SU(6)$  flavor group are shown in Fig. \ref{Baryons}.
As for the predictions for mesons in Fig. \ref{VM}, only confirmed PDG~\cite{Amsler:2008xx} states are shown.
The Roper state $N(1440)$ and the $N(1710)$ are well accounted for in this model as the first  and second radial
states. Likewise the $\Delta(1660)$ corresponds to the first radial state of the $\Delta$ family. The model is  successful in explaining the important parity degeneracy observed in the light baryon spectrum, such as the $L\! =\!2$, $N(1680)\!-\!N(1720)$ degenerate pair and the $L=2$, $\Delta(1905), \Delta(1910), \Delta(1920), \Delta(1950)$ states which are degenerate
within error bars. Parity degeneracy of baryons is also a property of the hard wall model, but radial states are not well described in this model.~\cite{deTeramond:2005su}

\section{Realization of Chiral  Invariance in Light Front Holography}

The proton eigenstate in light-front holography with massless quarks
\begin{equation}
\psi(\zeta) = \psi_+(\zeta) u_+  +  \psi_-(\zeta) u_-,
\end{equation}
 has $L^z=0$ and $L^z= + 1$ orbital components combined with spin components $S^z = + 1/2$ and $S^z = -1/2$ respectively. 
 The four-dimensional spinors $u_\pm$ are chiral spinors:
 $\gamma_5 u_\pm = \pm u_\pm$. The light-front Fock components $\psi_+$ and $\psi_-$ have  equal probability
 \begin{equation} \label{intpm}
 \int d \zeta\,  \vert \psi_+(\zeta) \vert^2 = \int d \zeta \, \vert \psi_-(\zeta) \vert^2,
 \end{equation}
 a manifestation of the  chiral invariance of the theory for massless quarks. As the equality expressed in (\ref{intpm}) follows from integrating the square of the wavefunction for all values of the holographic variable, {\it i.e.}, over all scales, it is a global property of the theory. On the other hand, for a given scale (a given value of the $z$ or $\zeta$) the solution is not chiral invariant as
plus an minus components have different twist, thus different $\zeta$ behavior. 
At the AdS boundary, $z \to 0$ or $\zeta \to 0$, the solution   is trivially  chiral symmetric since  $\psi_+ (\zeta) \to 0$  and  $\psi_- (\zeta) \to 0$ as $\zeta \to 0$. 
However, as we go to finite values of $z$ (or $\zeta$) -- measuring the proton at different scales -- both components plus and minus evolve differently, thus 
giving locally (for a given scale) chiral symmetry breaking effects, even for massless quarks. The measure of CSB depends in the scale $\kappa$ or $\Lambda_{\rm QCD}$.
Chiral symmetry invariance is thus manifest as a global property, and non-perturbative chiral symmetry breaking effects arise locally for AdS theories dual to confining gauge theories, even in the absence of quark masses.

\section{The Phenomenology of Exclusive Processes}

Exclusive processes play a key role in quantum chromodynamics, testing the primary quark and gluon interactions of QCD and the structure of hadrons at the amplitude level.  Two basic pictures have emerged based on perturbative QCD  (pQCD) and nonperturbative AdS/QCD. In pQCD
hard scattering amplitudes at a high scale $Q^2 >> \Lambda^2_{\rm QCD}$  factorize as a convolution of gauge-invariant hadron distribution amplitudes $\phi_H(x_i,Q)$ with the underlying hard scattering quark-gluon subprocess amplitude $T_H$.
The leading power fall-off of the hard scattering amplitude follows from the conformal scaling of
the underlying hard-scattering amplitude: $T_H \sim 1/Q^{n-4}$, where $n$ is
the total number of fields (quarks, leptons, or gauge fields)
participating in the hard
scattering.~\cite{Brodsky:1973kr, Matveev:1973ra} Thus the reaction
is dominated by subprocesses and Fock states involving the minimum
number of interacting fields.  In the case of $2 \to 2$ scattering
processes, this implies the dimensional counting rules $ {d\sigma/ dt}(A B \to C D) ={F_{A B \to C
D}(t/s)/ s^{n-2}},$ where $n = N_A + N_B + N_C +N_D$ and
$N_H$
is the minimum number of constituents of $H$.   The result is modified by the 
ERBL evolution~\cite{Lepage:1979zb,Efremov:1979qk} of the distribution amplitudes and the running of the QCD coupling

It is striking that the dimensional counting rules are also a key feature of nonperturbative AdS/QCD models~\cite{Polchinski:2001tt}.  Although the mechanisms are different, both the pQCD and AdS/QCD approaches depend on the leading twist interpolating operators of the hadron and their structure at short distances. In both theories, hadronic form factors at high $Q^2$ are dominated by the wavefunctions at small impact separation. 
This in turn leads to the color transparency phenomena.~\cite{Brodsky:1988xz,Bertsch:1981py}
For example, measurements of pion photoproduction
are consistent with dimensional counting $s^{7}{ d\sigma/dt}(\gamma p \to \pi^+  n) \sim $ constant
at fixed CM angle for $s > 7 $ GeV.   The angular distributions seen in hard  large CM angle scattering reactions are consistent with quark interchange,~\cite{Gunion:1973ex} a result also predicted by the hard wall AdS/QCD model.
Reviews are given in Refs.~\cite{Sivers:1975dg} and~\cite{White:1994tj}.
One sees the onset of perturbative QCD scaling behavior even for
exclusive nuclear amplitudes such as deuteron photodisintegration
(Here $n = 1+ 6 + 3 + 3 = 13 $)
and $s^{11}{ d\sigma/dt}(\gamma d \to p n) \sim $ constant at fixed CM
angle.~\cite{Holt:1990ze,Bochna:1998ca,Rossi:2004qm} The measured
deuteron form factor~\cite{Rock:1991jy} also appears to follow the
leading-twist QCD predictions~\cite{Brodsky:1976rz} at large
momentum transfers in the few GeV region.  The six color-triplet quarks of the valence Fock state of the deuteron can be arranged as a sum of five different color-singlet states, only one of which can be identified with the neutron-proton state and can account for the large magnitude of the deuteron form factor at high scales.
A measurement of ${d\sigma/dt}(\gamma d \to \Delta^{++}\Delta)$  in the scaling region
can establish the role of ``hidden-color" degrees of
freedom~\cite{Brodsky:1983vf} of the nuclear wavefunction in hard
deuteron reactions.

In the case of pQCD, the near-constancy of the
effective QCD coupling at small scales helps explain the general empirical
success of the dimensional counting rules for the near-conformal power
law fall-off of form factors and fixed-angle
scaling.~\cite{Brodsky:1989pv}

Color transparency~\cite{Brodsky:1988xz,Bertsch:1981py} is a key
property of color gauge theory, and it thus stands at the
foundations of QCD. Color transparency has been confirmed in
diffractive dijet production,~\cite{Aitala:2000hc} pion
photoproduction~\cite{Clasie:2007gq} and vector meson
electroproduction,~\cite{Airapetian:2002eh} but it is very important
to also systematically validate it in large-angle hadron scattering
processes.   Color transparency and higher-twist subprocesses
~\cite{Berger:1979du, Berger:1980qg, Berger:1981fr, Arleo:2010yg, Arleo:2009ch} where the trigger hadron is produced directly, such as $u u \to p \bar d,$ can account for the anomalous growth of the baryon-to-meson ratio with increasing centrality observed  in heavy ion collisions at RHIC.~\cite{Brodsky:2008qp}

\section{ Anomalies in Exclusive Processes}

Some exceptions to the general success of dimensional counting are known:

The transition form factor $F(Q^2)_{\gamma \to \pi^0}$ between a real photon and a pion has been  measured at BaBar to high $Q^2 \simeq 10~{\rm GeV}^2$, falling at high photon virtuality roughly as $1/ Q^{3/2}$ rather than the predicted $1/Q^2$ fall-off.  In contrast, preliminary measurements from BaBar~\cite{Druzhinin:2009gq} indicate that the transition form factors $F(Q^2)_{\gamma \to \eta}$
and $F(Q^2)_{\gamma \to \eta^\prime}$ are consistent  with the pQCD expectations. The photon to meson transition form factor is the simplest QCD hadronic exclusive amplitude, and thus it is critical to understand this discrepancy. As we shall discuss below, AdS/QCD predicts a broad distribution amplitude $\phi_\pi(x,Q) $ in the nonperturbative domain, but since ERBL evolution leads to a narrower distribution in the high $Q$ domain,  it cannot explain the BaBar anomaly.  It is difficult to imagine that the pion distribution amplitude is close to flat~\cite{Radyushkin:2009zg, Polyakov:2009je, Mikhailov:2009sa, Broniowski:2010yg} since this corresponds to a pointlike non-composite hadron.  It is  crucial to measure ${d\sigma\over dt }( \gamma \gamma \to  \pi^0 \pi^0)$  since the CM angular distribution is very sensitive to the shape of $\phi_\pi(x,Q)$.~\cite{Brodsky:1981rp}

The Hall A collaboration~\cite{Danagoulian:2007gs} at JLab
has reported another significant exception to the general empirical
success of dimensional counting in fixed-CM-angle Compton scattering
${d\sigma/dt}(\gamma p \to \gamma p) \sim {F(\theta_{CM})/s^8} $
instead of the predicted $1/s^6$ scaling.  The deviations from
fixed-angle conformal scaling may be due to corrections from
resonance contributions in the JLab energy range. It is interesting
that the hadron form factor $R_V(t)$,~\cite{Diehl:1998kh} which
multiplies the $\gamma q \to \gamma q$ amplitude is found by Hall A
to scale as $1/t^2$, in agreement with the  pQCD and AdS/QCD
prediction. In addition, the Belle measurement~\cite{Chen:2001sm} of
the timelike two-photon cross section ${d \sigma/dt}(\gamma \gamma
\to p \bar p)$ is consistent with $1/ s^6$ scaling.

Although large-angle proton-proton elastic scattering is well
described by dimensional scaling $s^{10}{ d\sigma/dt}(p p \to p p)
\sim $ constant at fixed CM angle,  extraordinarily  large spin-spin
correlations are observed.~\cite{Court:1986dh} The ratio of
scattering cross sections for spin-parallel and normal to the
scattering plane versus spin-antiparallel reaches $R_{NN} \simeq 4$
in large angle $ p p \to p p$ at $\sqrt s \simeq 5~$GeV; this is a
remarkable example of ``exclusive transversity".  Color transparency
is observed at lower energies but it fails~\cite{Mardor:1998zf} at
the same energy where $R_{NN}$ becomes large.  In fact, these
anomalies have a natural explanation~\cite{Brodsky:1987xw} as a
resonance effect related to the charm threshold in $pp$ scattering.
Alternative explanations of the large spin correlation are discussed
and reviewed in Ref.~\cite{Dutta:2004fw}.
Resonance formation is a natural phenomenon when all constituents
are relatively at rest. For example, a resonance effect can occur
due to the intermediate state  $uud uud c \bar c$ at the charm
threshold $\sqrt s = 5$ GeV in $p p$ collisions. Since the $c $ and
$\bar c$ have opposite intrinsic parity, the resonance appears in
the $L= J= S =1$ partial wave for $ p p \to p p$ which is only
allowed for spin-parallel and normal
to the  scattering plane
$A_{NN}=1$.~\cite{Brodsky:1987xw} Resonance formation at the charm
threshold also explains the dramatic quenching of color transparency
seen in quasielastic $p n$ scattering by the EVA BNL
experiment~\cite{Mardor:1998zf} in the same kinematic region. The
reason why these effects are apparent in $p p \to p p$ scattering
is that the amplitude for the formation of an $uud uud c \bar c$
$s$-channel resonance in the intermediate state is of the same
magnitude as the fast-falling background $p p \to  p p$ pQCD
amplitude from quark interchange at large CM angles: $M(p p \to pp)
\sim {1/u^2 t^2}$.   The open charm cross
section in $p p$ scattering  is predicted by unitarity to be of order of $1 ~\mu b$ at
threshold.~\cite{Brodsky:1987xw}  One also expects similar novel QCD phenomena in
large-angle photoproduction $\gamma p \to \pi N$ near the charm
threshold, including the breakdown of color transparency and strong
spin-spin correlations. These effects  can be tested by measurements
at the new JLab 12 GeV facility,  which would confirm resonance
formation in a low partial wave in $\gamma p \to \pi N$ at $\sqrt s
\simeq 4$ GeV due to attractive forces in the $uu d \bar c c$
channel.

Another difficulty for the application of pQCD to exclusive processes is the famous $J/\psi \to \rho \pi$ puzzle;  the observed unusually large branching ratio for $J/\psi \to \rho \pi$. In contrast, the branching ratio for $\Psi^\prime \to \rho \pi$ is very small. Such decays into pseudoscalar plus vector mesons require light-quark helicity suppression or internal orbital angular momentum and thus should be suppressed by hadron helicity conservation in pQCD.
However, the $J/\psi \to \rho \pi$  puzzle can be explained by the presence of intrinsic charm Fock states in the outgoing mesons.~\cite{Brodsky:1997fj}

\section{Nucleon Form Factor in Light-Front Holography}

\begin{figure}[h]
\begin{center}
 \includegraphics[width=6.9cm]{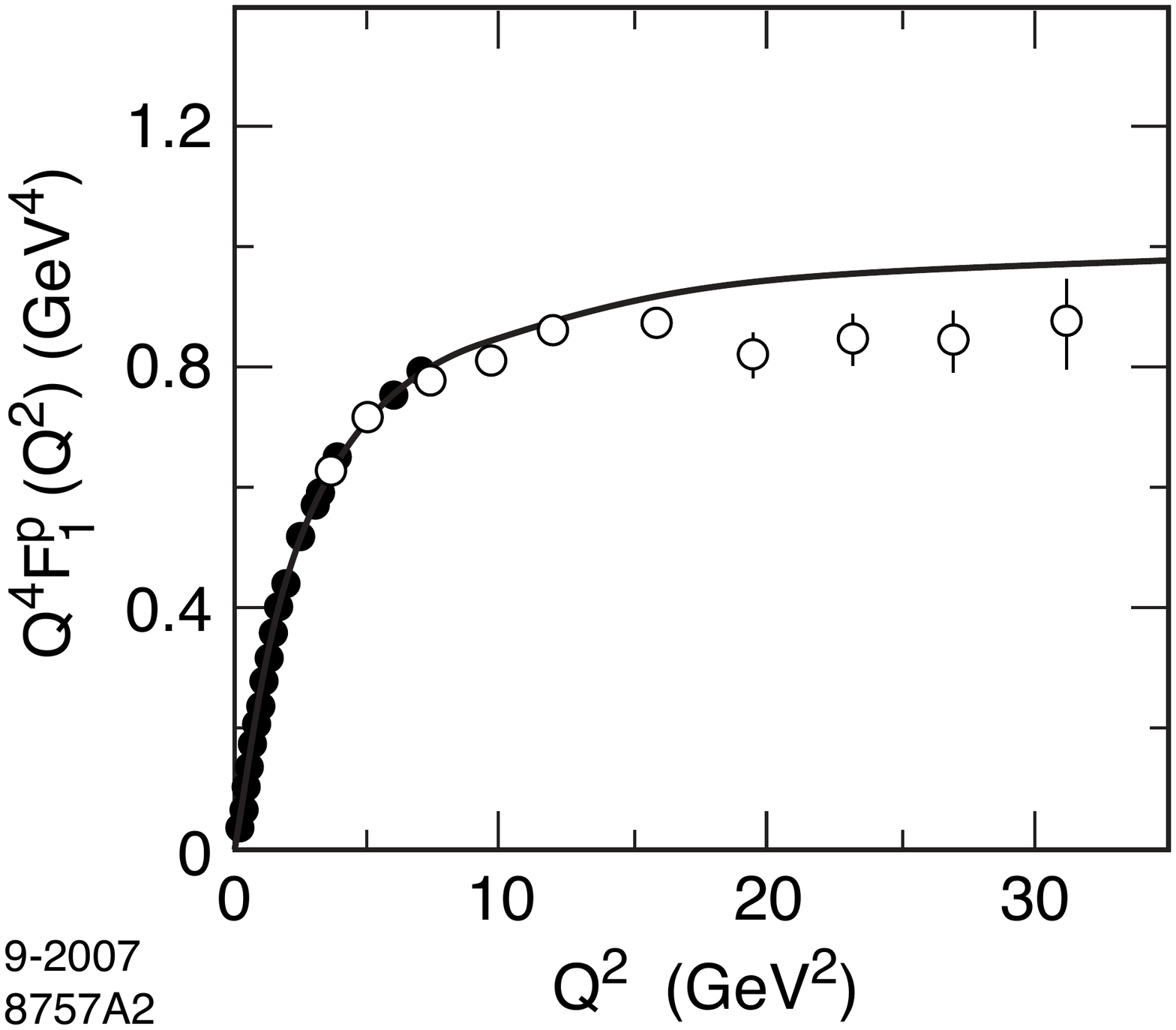}  \hspace{-42pt}  
\includegraphics[width=6.9cm]{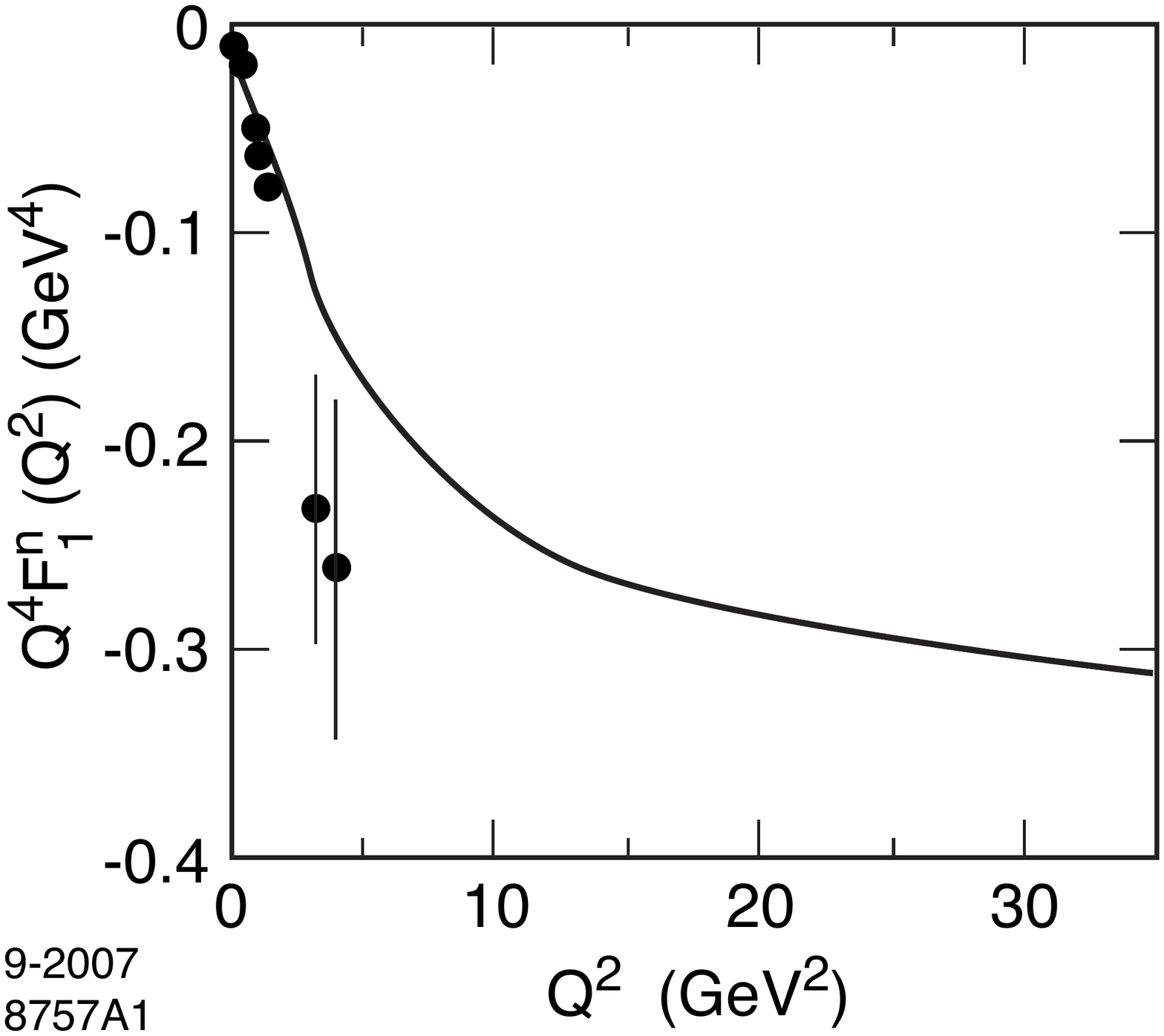}
 \caption{Predictions for $Q^4 F_1^p(Q^2)$ and $Q^4 F_1^n(Q^2)$ in the
soft wall model for $\kappa =  0.424$ GeV.~\cite{note3}}
\label{fig:nucleonFF}
\end{center}
\end{figure}

As an example  of the scaling behavior of a twist $\tau = 3$ hadron, we compute the spin non-flip
nucleon form factor in the soft wall model.~\cite{Brodsky:2008pg} The proton and neutron Dirac
form factors are given by
\begin{equation}
F_1^p(Q^2) =  \! \int  d \zeta \, J(Q, \zeta) \,
  \vert \psi_+(\zeta)\vert^2 ,
\end{equation}
\begin{equation}
F_1^n(Q^2) =  - \frac{1}{3}  \! \int  d \zeta  \,  J(Q, \zeta)
 \left[\vert \psi_+(\zeta)\vert^2 - \vert\psi_-(\zeta)\vert^2\right],
 \end{equation}
where $F_1^p(0) = 1$,~ $F_1^n(0) = 0$. The non-normalizable mode
 $J(Q,z)$ is the solution of the
AdS wave equation for the external electromagnetic current in presence of a dilaton
background  $\exp(\pm \kappa^2 z^2)$.~\cite{Brodsky:2007hb, Grigoryan:2007my}
Plus and minus components of the twist 3 nucleon LFWF are
\begin{equation} \label{eq:PhipiSW}
\psi_+(\zeta) \!=\! \sqrt{2} \kappa^2 \, \zeta^{3/2}  e^{-\kappa^2 \zeta^2/2},  ~~~
\Psi_-(\zeta) \!=\!  \kappa^3 \, \zeta^{5/2}  e^{-\kappa^2 \zeta^2/2}.
\end{equation}
The results for $Q^4 F_1^p(Q^2)$ and $Q^4 F_1^n(Q^2)$   and are shown in
Fig. \ref{fig:nucleonFF}.

\section{Nonperturbative Running Coupling from Light-Front Holography \label{alphaAdS}}

The concept of a running coupling $\alpha_s(Q^2)$  in QCD is usually restricted to the perturbative domain.  However, as in QED, it is useful to define the coupling as an analytic function valid over the full space-like and time-like domains.
The study of the non-Abelian QCD coupling at small momentum transfer is a complex problem because of  gluonic self-coupling and color confinement.

The definition of the running coupling in perturbative quantum field theory is scheme-dependent.  As discussed by Grunberg,~\cite{Grunberg}  an effective coupling or charge can be defined directly from physical observables.
Effective charges defined from
different observables can be related  to each other in the leading-twist domain using commensurate scale relations
 (CSR).~\cite{CSR}    The  potential between infinitely heavy quarks can be defined analytically in momentum transfer
 space as the product  of the running coupling times the Born gluon propagator: $V(q)  = - 4 \pi C_F {\alpha_V(q) / q^2}$.   This effective charge defines a renormalization scheme -- the $\alpha_V$ scheme of Appelquist, Dine,  and Muzinich.~\cite{Appelquist:1977tw}
In fact, the holographic coupling $\alpha_s^{AdS}(Q^2)$ can be considered to be the nonperturbative extension of the
$\alpha_V$ effective charge defined in Ref. \cite{Appelquist:1977tw}.
We can also make  use of the $g_1$ scheme, where the strong coupling $\alpha_{g_1}(Q^2)$ is determined from
the Bjorken sum rule.~\cite{BjorkenSR} The coupling $\alpha_{g_1}(Q^2)$ has the advantage that it is the best-measured effective charge, and it can be used to extrapolate the definition of the effective coupling to large distances.~\cite{Deur:2009hu}  Since $\alpha_{g_1}$ has been measured at intermediate energies, it is
particularly useful for studying  the transition from
small to large distances.

We have also   shown~\cite{Brodsky:2010ur}   how the LF holographic mapping of effective classical gravity in AdS space, modified by a positive-sign dilaton background, can  be used to identify an analytically simple  color-confining
non-perturbative effective coupling $\alpha_s^{AdS}(Q^2)$ as a function of the space-like momentum transfer $Q^2 = - q^2$.   This coupling incorporates  confinement
and agrees well with effective charge observables and lattice simulations.
It also exhibits an infrared fixed point at small $Q^2$ and asymptotic freedom at large $Q^2$. However, the fall-off   of
$\alpha_s^{AdS}(Q^2)$  at large $Q^2$ is exponential: $\alpha_s^{AdS}(Q^2) \sim e^{-Q^2 /  \kappa^2}$, rather than the
pQCD logarithmic fall-off.    It agrees with hadron physics data
extracted phenomenologically from different observables, as well as with  the predictions of models with built-in confinement  and lattice simulations. We
also show that a phenomenological extended coupling can be defined which implements the pQCD behavior.
The  $\beta$-function derived from light-front holography becomes significantly  negative in the non-perturbative regime $Q^2 \sim \kappa^2$, where it reaches a minimum, signaling the transition region from the infrared (IR) conformal region, characterized by hadronic degrees of freedom,  to a pQCD conformal ultraviolet (UV)  regime where the relevant degrees of freedom are the quark and gluon constituents.  The  $\beta$-function vanishes at large $Q^2$ consistent with asymptotic freedom, and it vanishes at small $Q^2$ consistent with an infrared fixed point.~\cite{Brodsky:2008be, Cornwall:1981zr}

Let us consider a five-dimensional gauge field $F$ propagating in AdS$_5$ space in presence of a dilaton background
$\varphi(z)$ which introduces the energy scale $\kappa$ in the five-dimensional action.
At quadratic order in the field strength the action is
\begin{equation}
S =  - {1\over 4}\int \! d^5x \, \sqrt{g} \, e^{\varphi(z)}  {1\over g^2_5} \, F^2,
\label{eq:action}
\end{equation}
where the metric determinant of AdS$_5$ is $\sqrt g = ( {R/z})^5$,  $\varphi=  \kappa^2 z^2$ and the square of the coupling $g_5$ has dimensions of length.   We  can identify the prefactor
\begin{equation} \label{eq:flow}
g^{-2}_5(z) =  e^{\varphi(z)}  g^{-2}_5 ,
\end{equation}
in the  AdS  action (\ref{eq:action})  as the effective coupling of the theory at the length scale $z$.
The coupling $g_5(z)$ then incorporates the non-conformal dynamics of confinement. The five-dimensional coupling $g_5(z)$
is mapped,  modulo a  constant, into the Yang-Mills (YM) coupling $g_{YM}$ of the confining theory in physical space-time using light-front holography. One  identifies $z$ with the invariant impact separation variable $\zeta$ which appears in the LF Hamiltonian:
$g_5(z) \to g_{YM}(\zeta)$. Thus
\begin{equation}  \label{eq:gYM}
\alpha_s^{AdS}(\zeta) = g_{YM}^2(\zeta)/4 \pi \propto  e^{-\kappa^2 \zeta^2} .
\end{equation}

In contrast with the 3-dimensional radial coordinates of the non-relativistic Schr\"o-dinger theory, the natural light-front
variables are the two-dimensional cylindrical coordinates $(\zeta, \phi)$ and the
light-cone fraction $x$. The physical coupling measured at the scale $Q$ is the two-dimensional Fourier transform
of the  LF transverse coupling $\alpha_s^{AdS}(\zeta)$  (\ref{eq:gYM}). Integration over the azimuthal angle
 $\phi$ gives the Bessel transform
 \begin{equation} \label{eq:2dimFT}
\alpha_s^{AdS}(Q^2) \sim \int^\infty_0 \! \zeta d\zeta \,  J_0(\zeta Q) \, \alpha_s^{AdS}(\zeta),
\end{equation}
in the $q^+ = 0$ light-front frame where $Q^2 = -q^2 = - \mbf{q}_\perp^2 > 0$ is the square of the space-like
four-momentum transferred to the
hadronic bound state.   Using this ansatz we then have from  Eq.  (\ref{eq:2dimFT})
\begin{equation}
\label{eq:alphaAdS}
\alpha_s^{AdS}(Q^2) = \alpha_s^{AdS}(0) \, e^{- Q^2 /4 \kappa^2}.
\end{equation}
In contrast, the negative dilaton solution $\varphi=  -\kappa^2 z^2$ leads to an integral which diverges at large $\zeta$.
We identify $\alpha_s^{AdS}(Q^2)$ with the physical QCD running coupling
in its nonperturbative domain.

The flow equation  (\ref{eq:flow}) from the scale dependent measure for the gauge fields can be understood as a consequence of field-strength renormalization.
In physical QCD we can rescale the non-Abelian gluon field  $A^\mu \to \lambda A^\mu$  and field strength
$G^{\mu \nu}  \to \lambda G^{\mu \nu}$  in the QCD Lagrangian density  $\mathcal{L}_{\rm QCD}$ by a compensating rescaling of the coupling strength $g \to \lambda^{-1} g.$  The renormalization of the coupling $g _{phys} = Z^{1/2}_3  g_0,$   where $g_0$ is the bare coupling in the Lagrangian in the UV-regulated theory,  is thus  equivalent to the renormalization of the vector potential and field strength: $A^\mu_{ren} =  Z_3^{-1/2} A^\mu_0$, $G^{\mu \nu}_{ren} =  Z_3^{-1/2} G^{\mu \nu}_0$   with a rescaled Lagrangian density
${\cal L}_{\rm QCD}^{ren}  = Z_3^{-1}  { \cal L}_{\rm QCD}^0  = (g_{phys}/g_0)^{-2}  \mathcal{L}_0$.
 In lattice gauge theory,  the lattice spacing $a$ serves as the UV regulator, and the renormalized QCD coupling is determined  from the normalization of the gluon field strength as it appears  in the gluon propagator. The inverse of the lattice size $L$ sets the mass scale of the resulting running coupling.
As is the case in lattice gauge theory, color confinement in AdS/QCD reflects nonperturbative dynamics at large distances. The QCD couplings defined from lattice gauge theory and the soft wall holographic model are thus similar in concept, and both schemes are expected to have similar properties in the nonperturbative domain, up to a rescaling of their respective momentum scales.

\subsection{Comparison of  the Holographic Coupling with Other Effective Charges \label{alphatest}}

\begin{figure}[h]
\begin{center}
\includegraphics[width=7.0cm]{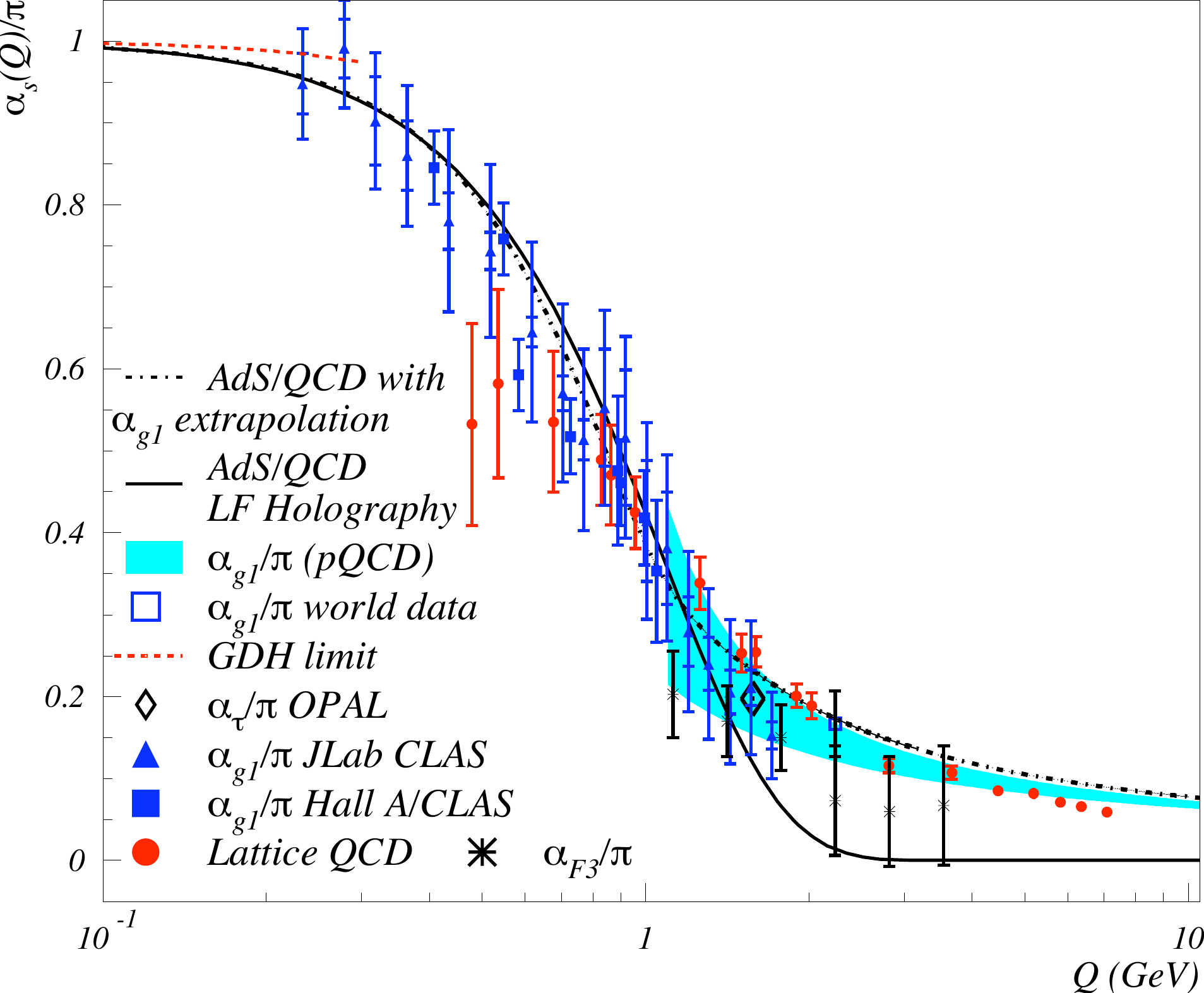} ~~~
\includegraphics[width=6.8cm]{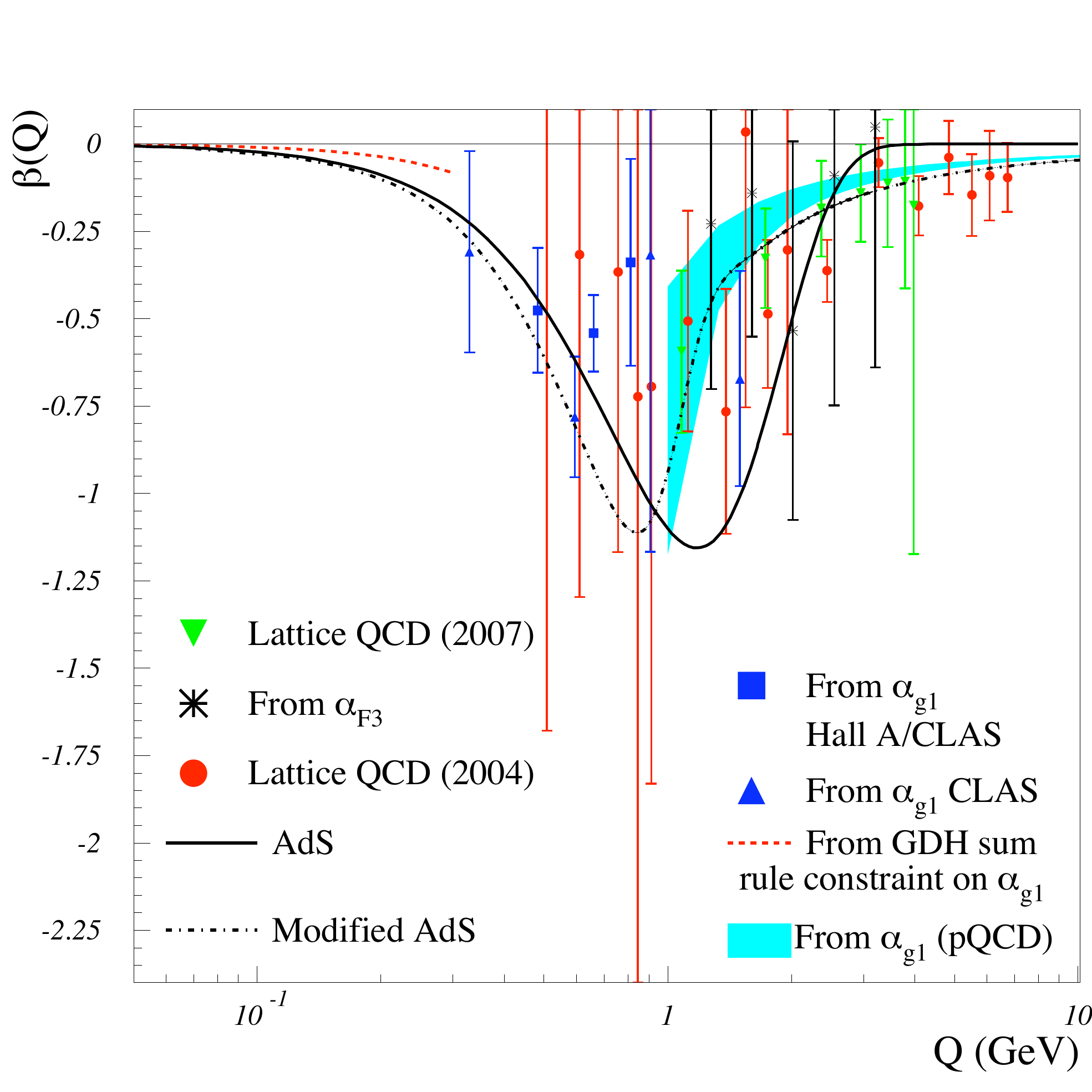}
 \caption{(a) Effective coupling  from LF holography  for  $\kappa = 0.54 ~ {\rm GeV}$ compared with effective QCD couplings  extracted from
different observables and lattice results. (b) Prediction for the $\beta$-function compared to  lattice simulations,  JLab and CCFR results  for the Bjorken sum rule effective charge.}
\label{alphas}
\end{center}
\end{figure}

The effective coupling  $\alpha^{AdS}(Q^2)$ (solid line) is compared in Fig. \ref{alphas} with  experimental and lattice data. For this comparison to be meaningful, we have to impose the same normalization on the AdS coupling as the $g_1$ coupling. This defines $\alpha_s^{AdS}$ normalized to the $g_1$ scheme: $\alpha_{g_1}^{AdS}\left(Q^2 \! =0\right) = \pi.$
Details on the comparison with other effective charges are given in Ref. ~\cite{Deur:2005cf}.

The couplings in Fig. \ref{alphas} (a) agree well in the strong coupling regime  up to $Q  \! \sim \! 1$ GeV.  The value $\kappa  = 0.54 ~ {\rm GeV}$ is determined from the vector meson Regge trajectory.~\cite {deTeramond:2009xk}
The lattice results shown in Fig. \ref{alphas} from Ref.~\cite{Furui:2006py} have been scaled to match the perturbative UV domain. The effective charge $\alpha_{ g_1}$ has been determined in Ref.~\cite{Deur:2005cf} from several experiments.  Fig. \ref{alphas} also displays other couplings from different observables as well as $\alpha_{g_1}$ which is computed from the
Bjorken sum rule~\cite{BjorkenSR}  over a large range of momentum transfer (cyan band). 
A recent analysis~\cite{Aguilar:2009nf} using the pinch scheme~\cite{Cornwall:1981zr} predicts a similar fixed-point behavior.

At $Q^2\!=\!0$ one has the constraint  on the slope of $\alpha_{g_1}$ from the Gerasimov-Drell-Hearn (GDH) sum rule~\cite{GDH} which is also shown in the figure.
The results show no sign of a phase transition, cusp, or other non-analytical behavior, a fact which allows us to extend the functional dependence of the coupling to large distances.
As discussed below,
the smooth behavior of the  AdS strong
coupling also allows us to extrapolate its form to the perturbative domain.~\cite{Brodsky:2010ur}

The hadronic model obtained from the dilaton-modified AdS space provides a semi-classical first approximation to QCD.  Color confinement is introduced by the harmonic oscillator potential, but effects from gluon creation and absorption are not included in this effective theory.  The nonperturbative  confining effects vanish exponentially at large momentum transfer (Eq. (\ref{eq:alphaAdS})), and thus the logarithmic fall-off from pQCD quantum loops will dominate in this regime.
The  running coupling  $\alpha_s^{AdS}$ given by Eq.  (\ref{eq:alphaAdS})  is obtained from a
color-confining potential.   Since the strong coupling is an analytical function of the momentum transfer at all scales, we can extend  the range of applicability of $\alpha_s^{AdS}$ by matching to a  perturbative coupling
at the transition scale  $Q \sim 1$ GeV, where pQCD contributions become important, as described in Ref. \cite{Brodsky:2010ur}.
The smoothly extrapolated result (dot-dashed line) for  $\alpha_{s}$ is also shown on Fig.~\ref{alphas}.
In order to have a fully analytical model, we write
\begin{equation}
\label{eq:alphafit}
\alpha_{Modified, g_1}^{AdS}(Q^2) = \alpha_{g_1}^{AdS}(Q^2) g_+(Q^2 ) + \alpha_{g_1}^{fit}(Q^2) g_-(Q^2),
\end{equation}
where $g_{\pm}(Q^2) = 1/(1+e^{\pm \left(Q^2 - Q^2_0\right)/\tau^2})$ are smeared step functions which match the two
regimes. The parameter $\tau$ represents the width of the
transition region.    Here $\alpha_{g_1}^{AdS}$ is given by Eq. (\ref{eq:alphaAdS}) with the normalization  $\alpha_{g_1}^{AdS}(0)=\pi$
-- the plain black line in Fig.~\ref{alphas} -- and $\alpha_{g_{1}}^{fit}$ in Eq. (\ref{eq:alphafit}) is the analytical
fit to the measured coupling $\alpha_{g_1}$.~\cite{Deur:2005cf}
The couplings  are chosen to have the same  normalization at $Q^2=0.$
The smoothly extrapolated result (dot-dashed line) for  $\alpha_{s}$ is also shown on Fig.~\ref{alphas}. We use the
parameters $Q_{0}^{2}=0.8$ GeV$^{2}$ and $\tau^2=0.3$ GeV$^{2}$.

\subsection{The $\beta$-Function from AdS/QCD}

The $\beta$-function for the nonperturbative  effective coupling obtained from the LF holographic mapping in a positive dilaton modified AdS background  is
\begin{equation} \label{eq:beta}
\beta^{AdS}(Q^2)  = {d \over d \log{Q^2}}\alpha^{AdS}(Q^2) = - {\pi Q^2\over 4 \kappa^2} e^{-Q^2/(4 \kappa^2)}.
\end{equation}
The solid line in Fig. \ref{alphas} (b) corresponds to the light-front holographic result Eq.  (\ref{eq:beta}).    Near $Q_0 \simeq 2 \kappa \simeq 1$ GeV, we can interpret the results as a transition from  the nonperturbative IR domain to the quark and gluon degrees of freedom in the perturbative UV  regime. The transition momentum  scale $Q_0$  is compatible with the momentum transfer for the onset of scaling behavior in exclusive reactions where quark counting rules are observed.~\cite{Brodsky:1973kr}
For example, in deuteron photo-disintegration the onset of scaling corresponds to  momentum transfer  of  1.0  GeV to the nucleon involved.~\cite{Gao:2004zh}  Dimensional counting is built into the AdS/QCD soft and hard wall models since the AdS amplitudes $\Phi(z)$ are governed by their twist scaling behavior $z^\tau$ at short distances, $ z \to 0$.~\cite{Polchinski:2001tt}

Also shown on Fig. \ref{alphas} (b) are the $\beta$-functions obtained from phenomenology and lattice calculations. For clarity, we present only the LF holographic predictions, the lattice results from,~\cite{Furui:2006py} and the
experimental data supplemented by the relevant sum rules.
The dot-dashed curve corresponds to the
extrapolated approximation obtained by matching to AdS results to the perturbative coupling~\cite{Brodsky:2010ur}
given by Eq. (\ref{eq:alphafit}).
The $\beta$-function extracted from LF holography, as well as the forms obtained from
the works of Cornwall,~\cite{Cornwall:1981zr}  Bloch, Fisher {\it et al.},~\cite{S-Deq.} Burkert and Ioffe~\cite{Burkert-Ioffe} and Furui and Nakajima,~\cite{Furui:2006py} are seen to have a similar shape and magnitude.

Judging from these results, we infer that the   actual  $\beta$-function of QCD will extrapolate between the non-perturbative results for $Q < 1$ GeV and the pQCD results
for $Q > 1$ GeV. We also observe that the general conditions
\begin{eqnarray}
& \beta(Q \to 0) =  \beta(Q \to \infty) = 0 , \label{a} \\
&  \beta(Q)  <  0, ~ {\rm for} ~  Q > 0 , \label{b}\\
& \frac{d \beta}{d Q} \big \vert_{Q = Q_0}  = 0, \label{c} \\
& \frac{d \beta}{d Q}   < 0, ~ {\rm for} ~ Q < Q_0, ~~
 \frac{d \beta}{d Q}   > 0, ~ {\rm for} ~ Q > Q_0 \label{d} .
\end{eqnarray}
are satisfied by our model $\beta$-function obtained from LF holography.

Eq. (\ref{a}) expresses the fact that  QCD approaches a conformal theory in both the far ultraviolet and deep infrared regions. In the semiclassical approximation to QCD  without particle creation or absorption,
the $\beta$-function is zero and the approximate theory is scale  invariant
in the limit of massless quarks.~\cite{Parisi:1972zy} When quantum corrections are included,
the conformal behavior is
preserved at very large $Q$ because of asymptotic freedom and near $Q \to 0$ because the theory develops a fixed
point.  An infrared fixed point is in fact a natural consequence of color confinement:~\cite{Cornwall:1981zr}
since the propagators of the colored fields have a maximum wavelength,  all loop
integrals in the computation of  the gluon self-energy decouple at $Q^2 \to 0$.~\cite{Brodsky:2008be} Condition (\ref{b}) for $Q^2$ large, expresses the basic anti-screening behavior of QCD where the strong coupling vanishes. The $\beta$-function in QCD is essentially negative, thus the coupling increases monotonically from the UV to the IR where it reaches its maximum value:  it has a finite value for a theory with a mass gap. Equation (\ref{c}) defines the transition region at $Q_0$ where the 
$\beta$-function has a minimum.  Since there is only one hadronic-partonic transition, the minimum is an absolute minimum; thus the additional conditions expressed in Eq (\ref{d}) follow immediately from
Eqs.  (\ref{a}-\ref{c}). The conditions given by Eqs.  (\ref{a}-\ref{d}) describe the essential
behavior of the full $\beta$-function for an effective QCD coupling whose scheme/definition is similar to that of the $V$-scheme.

\section {Vacuum Effects and Light-Front Quantization}

The LF vacuum is remarkably simple in light-cone quantization because of the restriction $k^+ \ge 0.$   For example in QED,  vacuum graphs such as $e^+ e^- \gamma $  associated with the zero-point energy do not arise. In the Higgs theory, the usual Higgs vacuum expectation value is replaced with a $k^+=0$ zero mode;~\cite{Srivastava:2002mw} however, the resulting phenomenology is identical to the standard analysis.

Hadronic condensates play an important role in quantum chromodynamics.
It is widely held that quark and gluon vacuum condensates have a physical existence, independent of  hadrons, measurable spacetime-independent configurations of QCD's elementary degrees-of-freedom in a hadron-less ground state. However, a  non-zero spacetime-independent QCD vacuum condensate poses a critical dilemma for gravitational interactions because it would lead to a cosmological constant some 45 orders of magnitude larger than observation.  As noted  in Ref. \cite{Brodsky:2009zd}, this conflict is avoided if strong interaction condensates are properties of the light-front wavefunctions of the hadrons, rather than the hadron-less ground state of QCD.

The usual assumption that non-zero vacuum condensates exist and possess a measurable reality has long been recognized as posing a conundrum for the light-front formulation of QCD.   In the light-front formulation, the ground-state is a structureless Fock space vacuum, in which case it would seem to follow that dynamical chiral symmetry breaking (CSB)  is impossible.  In fact,  as first argued by Casher and Susskind \cite{Casher:1974xd} dynamical CSB must be a property of hadron wavefunctions, not of the vacuum  in the light-front framework.  This thesis has also been explored in a series of recent articles
\cite{Brodsky:2008tk, Brodsky:2008be, Brodsky:2009zd}.

Conventionally, the quark and gluon condensates are considered to be properties
of the QCD vacuum and hence to be constant throughout spacetime.
A new perspective on the nature of QCD
condensates $\langle \bar q q \rangle$ and $\langle
G_{\mu\nu}G^{\mu\nu}\rangle$, particularly where they have spatial and temporal
support,
has recently been presented.~\cite{Brodsky:2008be,Brodsky:2009zd,Brodsky:2008xm,Brodsky:2008xu}
Their spatial support is restricted to the interior
of hadrons, since these condensates arise due to the interactions of quarks and
gluons which are confined within hadrons.

For example, consider a meson consisting of a light quark $q$ bound to a heavy
antiquark, such as a $B$ meson.  One can analyze the propagation of the light
$q$ in the background field of the heavy $\bar b$ quark.  Solving the
Dyson-Schwinger equation for the light quark one obtains a nonzero dynamical
mass and, via the connection mentioned above, hence a nonzero value of the
condensate $\langle \bar q q \rangle$.  But this is not a true vacuum
expectation value; instead, it is the matrix element of the operator $\bar q q$
in the background field of the $\bar b$ quark.  The change in the (dynamical)
mass of the light quark in this bound state is somewhat reminiscent of the
energy shift of an electron in the Lamb shift, in that both are consequences of
the fermion being in a bound state rather than propagating freely.
Similarly, it is important to use the equations of motion for confined quarks
and gluon fields when analyzing current correlators in QCD, not free
propagators, as has often been done in traditional analyses of operator
products. Since the distance between the
quark and antiquark cannot become arbitrarily large, one cannot create a quark
condensate which has uniform extent throughout the universe.  
Thus in a fully self-consistent treatment of the bound state, this phenomenon occurs in the background field of the $\bar b$-quark, whose influence on light-quark propagation is primarily concentrated in the far infrared and whose presence ensures the manifestations of light-quark dressing are gauge invariant.

In the case of the pion one finds that the vacuum quark condensate  that appears in the  Gell Mann-Oakes Renner formula, is, in fact, a chiral-limit value of an `in-pion' condensate. This condensate is no more a property of the ``vacuum'' than the pion's chiral-limit leptonic decay constant.  
One can connect the Bethe-Salpeter formalism to the light-front formalism, by fixing the light-front time $\tau$. This then leads to the Fock state expansion.  In fact,  dynamical CSB in the light-front formulation, expressed via `in-hadron' condensates, can be shown to be connected with sea-quarks derived from higher Fock states.  This solution is similar  to that discussed in Ref.\,\cite{Casher:1974xd}. 
Moreover, Ref. \cite{Langfeld:2003ye} establishes the equivalence of all three definitions of the vacuum quark condensate: a constant in the operator product expansion,~\cite{Lane:1974he,Politzer:1976tv} via the Banks-Casher formula,~\cite{Banks:1979yr} and the trace of the chiral-limit dressed-quark propagator.

AdS/QCD also provides  a description of chiral symmetry breaking by
using the propagation of a scalar field $X(z)$ which encodes the coupling
of the quark mass $m_q$ and quark condensate  $\langle \bar q q \rangle$ to the hadron. In the hard-wall model the AdS
solution has the form~\cite{Erlich:2005qh,DaRold:2005zs} 
\begin{equation}
X(z) = a_1 z+ a_2 z^3,
\end{equation}
 where $a_1$ is
proportional to the current-quark mass and  $a_2$ to the quark condensate $\langle \bar q q \rangle$.
Since the quark is a color nonsinglet, the propagation of $X(z),$ and thus the
domain of the quark condensate, is limited to the region of color confinement.
The effect of the $a_2$ term
varies within the hadron. This is consistent with the idea that the effect of CSB increases as one goes toward large interquark separation.  In the light-front view,  this is due to the higher Fock states.  In fact, that the CSB dynamics depends on the total size of the hadron, means that there will  be  unexpected dependence of the effective quark mass on the spectators.   For example, the  effect of the $u$  quark mass term will be smaller  
in a $B^+(\bar b u)$ than a $K^+(\bar s u)$ since the $B$ has a smaller transverse size.~\cite{Karliner:2008in}

The  picture of condensates with spatial support restricted to hadrons
is also in general agreement with results from chiral bag 
models,~\cite{Chodos:1975ix,Brown:1979ui,Hosaka:1996ee}
which modify the original MIT bag by coupling a pion field to the surface of
the bag in a chirally invariant manner.  It is important to notice however, that AdS/QCD   does not help to elucidate the nature of the in-hadron versus vacuum condensates.
The quark mass and the condensate are from the point of view of holography, at least in the fixed background approximation, arbitrary constants given by the initial conditions.
What is physically relevant from the AdS/QCD perspective is the coupling of $m_q$ and $\langle \bar q q \rangle$ to  $X(z)$, and this leads to observable effects.

\section{Conclusions \label{conclusions}}

The combination of  anti-de Sitter  space 
with light-front  holography provides an accurate first approximation for the spectra and wavefunctions of meson and baryon light-quark  bound states. 
This new framework provides an elegant connection between a semiclassical first approximation to QCD, quantized on the light-front, with hadronic modes propagating on a fixed AdS background. The resulting bound-state Hamiltonian equation of motion in QCD leads to  relativistic light-front wave equations in the invariant impact variable $\zeta$, which measures the separation of the quark and gluonic constituents within the hadron at equal light-front time. This corresponds
to the effective single-variable relativistic Schr\"odinger-like equation in the AdS fifth dimension coordinate $z$,  Eq. (\ref{eq:QCDLFWE}). The eigenvalues give the hadronic spectrum, and the eigenmodes represent the probability distributions of the hadronic constituents at a given scale.
In particular, the light-front holographic mapping of  effective classical gravity in AdS space, modified by a positive-sign dilaton background,  provides a very good description of the spectrum and form factors of light mesons and baryons.
We emphasize that 
the hadron eigenstate generally has components with different orbital angular momentum.  For example, the proton eigenstate in AdS/QCD with massless quarks has $L=0$ and $L=1$ light-front Fock components with equal probability -- a novel manifestation of of chiral invariance.

We have also shown that the light-front holographic mapping of  effective classical gravity in AdS space, modified by the positive-sign dilaton background  predicts the form of a non-perturbative effective coupling $\alpha_s^{AdS}(Q)$ and its $\beta$-function.
The AdS/QCD running coupling is in very good agreement with the effective
coupling $\alpha_{g_1}$ extracted from  the Bjorken sum rule.
The holographic $\beta$-function displays a transition from  nonperturbative to perturbative  regimes  at a momentum scale $Q \sim 1$ GeV.
Our analysis indicates that light-front holography captures the characteristics of the full  $\beta$-function of QCD  and  the essential dynamics of confinement, thus giving further support to the application of the gauge/gravity duality to the confining dynamics of strongly coupled QCD.

There are many phenomenological applications where detailed knowledge of the QCD coupling and the renormalized gluon propagator at relatively soft momentum transfer are essential.
This includes exclusive and semi-exclusive processes as well as the rescattering interactions which
create the leading-twist Sivers single-spin correlations in
semi-inclusive deep inelastic scattering,~\cite{Brodsky:2002cx, Collins:2002kn} the Boer-Mulders functions which lead to anomalous  $\cos 2 \phi$  contributions to the lepton pair angular distribution in the unpolarized Drell-Yan reaction,~\cite{Boer:2002ju}  and the Sommerfeld-Sakharov-Schwinger correction to heavy quark production at
threshold.~\cite{Brodsky:1995ds}
The confining AdS/QCD coupling from light-front holography thus can provide a
quantitative understanding of this factorization-breaking physics.~\cite{Collins:2007nk}

A new perspective on the nature of quark and gluon condensates in
quantum chromodynamics has also been addressed:~\cite{Brodsky:2008be,Brodsky:2008xm,Brodsky:2008xu}  the spatial support of QCD condensates
is restricted to the interior of hadrons, since they arise due to the
interactions of confined quarks and gluons.  In  LF theory, the condensate physics is replaced by the dynamics of higher non-valence Fock states as shown by Casher and Susskind.~\cite{Casher:1974xd}  In particular, chiral symmetry is broken in a limited domain of size $1/ m_\pi$,  in analogy to the limited physical extent of superconductor phases.  This novel description  of chiral symmetry breaking  in terms of ``in-hadron condensates"  has also been observed in Bethe-Salpeter studies.~\cite{Maris:1997hd,Maris:1997tm}
This picture also explains the
results of recent studies~\cite{Ioffe:2002be,Davier:2007ym,Davier:2008sk} which find no significant signal for the vacuum gluon
condensate.

\hspace{60pt}

\noindent{\bf Acknowledgments}

\vspace{5pt}

We thank  Alexander Deur, Craig Roberts, Robert Shrock, and Peter Tandy for  collaborations.
This research was supported by the Department of Energy contract DE--AC02--76SF00515.
Invited talk, presented by SJB at the 50th Crakow School, Zakopane, Poland in honor of its founder, Andre Bialas.
G. F. d. T. thanks the members of the High Energy Physics Group at Imperial College in London for their hospitality.

\end{document}